\documentclass[aps,prb,twocolumn,superscriptaddress]{revtex4-2}
\usepackage{amsmath,bm}
\usepackage{graphicx,float}
\usepackage[dvipsnames,svgnames,x11names]{xcolor}
\usepackage{microtype}
\usepackage{tabularx}
\usepackage{xr-hyper}
\usepackage{hyperref}

\hypersetup{
    colorlinks=true,
    citecolor=Fuchsia,
    urlcolor=Blue4,
    linkcolor=TealBlue,
}

\setcitestyle{super}
\makeatletter

\renewcommand\frontmatter@abstractwidth{\dimexpr\textwidth\relax}
\makeatother

\makeatletter
\let\old@bibcite\bibcite
\renewcommand{\bibcite}[2]{}
\makeatother
\makeatletter
\let\bibcite\old@bibcite
\makeatother

\newcommand{\supp}{\textbf{Supporting Information}}

\begin{document}

\title{Steady--State Current Signatures of Strong Light--Matter Coupling in Single--Molecule Junctions}
\author{Kritanjan Polley}
\affiliation{Simons Center for Computational Physical Chemistry at New York University, New York, New York 10003, USA}
\author{Norah M. Hoffmann}
\affiliation{Simons Center for Computational Physical Chemistry at New York University, New York, New York 10003, USA}
\affiliation{Department of Chemistry, New York University, New York, New York 10003, USA}
\affiliation{Department of Physics, New York University, New York, New York 10003, USA}
\email{nmh6061@nyu.edu}
\date{\today}

\begin{abstract}
 Strong light-matter coupling offers exciting opportunities for controlling molecular properties, yet resolving single-molecule behavior from collective effects remains a challenge. Recent experiments in scanning tunneling microscope break junctions (STM-BJs) have demonstrated strong light-matter coupling at the single-molecule level. Because STM-BJs provide direct access to molecular-junction currents, we investigate whether strong coupling leaves a measurable fingerprint in transport, potentially probing polaritonic states beyond optical spectroscopy. Using a semiclassical mapping approach for nonequilibrium quantum transport, we find that steady-state current indeed carries a direct signature of single-molecule strong coupling. We test the robustness of this signature against electrode connectivity, nuclear motion, spatially structured electromagnetic modes, and solvent effects. While environmental factors and nuclear motion reshape the signature, the current is not a passive response but is actively influenced by coupled molecular dynamics. Our work establishes current as a tool to interpret and control strong light–matter interactions in nanoscale junctions.
\end{abstract}

\maketitle

\begin{figure}[!htb]
    \centering
    \includegraphics[width=0.75\linewidth]{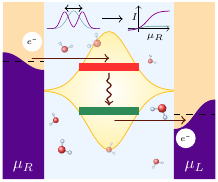}
    \par\medskip\textbf{TOC Graphic}
\end{figure}

Strong light-matter coupling arises from the hybridization of molecular excitations with highly confined electromagnetic modes, creating polaritonic states with the potential to reshape chemical energy landscapes,~\cite{thomas16,xiang24} material properties,~\cite{low17,naaman19} charge transport,~\cite{xiang16,su16,dief23,kamenetska10,tang23} and collective phenomena such as condensation.~\cite{Galbiati12} However, conventional strong light-matter coupling platforms commonly rely on collective coupling, where many emitters interact simultaneously with a cavity mode.~\cite{cui22} This collective nature makes it difficult to resolve the individual microscopic contributions of molecules or to map single-molecule dynamics onto the macroscopic system. 

One alternative route to strong light-matter coupling is offered by plasmonic cavities, which can achieve strong coupling at the single- to few-molecule limit by confining electromagnetic fields to deeply subwavelength volumes, although the metallic environment introduces strong plasmonic and excitonic losses.~\cite{baumberg19,wang22,hu24} Recent experimental observations in scanning tunneling microscope break junctions (STM-BJs) reveal electroluminescence with split peaks characteristic of strong light-matter coupling.~\cite{chikkaraddy16,benz16,martin20,kuisma22,bitton22,paoletta24} In such a junction, a single molecule bridges two metallic electrodes, simultaneously forming a conductive molecular junction and a tightly confined cavity.~\cite{muniain24,zhang15,zhao26} An important distinction of this platform from conventional plasmonic cavities is that these bias-induced polaritonic junctions are voltage-driven rather than laser-driven.~\cite{dolezal24,zheng25} The applied bias can generate the relevant interfacial excitation through resonant transport and, as shown in Ref.~\citenum{wang26}, can simultaneously reduce its coupling to metallic loss channels and increase the lifetime of the excitation.

This single-molecule platform extends the STM-BJ toolbox to strong light-matter coupling. STM-BJs already provide access to chemically specific conductance measurements,~\cite{Meisner12} quantum transport,~\cite{york25} and hence single-molecule control.~\cite{su16} Importantly, they also provide direct access to the current, which is highly sensitive to the vibronic structure and transport pathways of the molecular junction.~\cite{gao24,prana24} This raises a natural question: can strong light-matter coupling itself leave a measurable signature in the current? Strong coupling modifies the available pathways for charge transfer. If these changes are reflected directly in the junction current and conductance, transport measurements could provide a complementary route to probe strong light-matter coupling beyond conventional optical spectroscopy.~\cite{york25}

To this end, we utilize a semiclassical mapping framework, benchmarked against hierarchical equations of motion (HEOM)~\cite{tanimura20} in limiting cases, to investigate the steady-state current under varying strong light-matter coupling conditions. We systematically examine different molecule–electrode connectivities, the role of nuclear degrees of freedom, spatially structured light-matter coupling, and coupling to a solvent environment. Indeed, we find that strong light-matter coupling can produce a clear and coupling-dependent signature in the steady-state current. Nuclear motion and spatial variations of the electromagnetic field reshape this signature, while direct electrode-molecule transport pathways or sufficiently strong environmental coupling can substantially reduce or even suppress it. These results show that the current has the potential to provide a complementary probe to optical spectroscopy.

Modeling dynamics in single-molecule junctions is challenging due to their intrinsic nonequilibrium nature and the simultaneous interplay of electronic transport, photonic degrees of freedom, nuclear motion, and environmental effects.~\cite{thoss18} Numerically exact approaches such as HEOM become computationally demanding once complex nuclear dynamics and strong light-matter coupling are included.~\cite{jin08,haertle13,huang23} To bridge this gap, we combine semiclassical mapping frameworks for bosons~\cite{Meyer79a,runeson19,polley20} and fermions~\cite{montoya18,sun21} to treat all these components within a unified dynamical description. Specifically, we consider a minimal molecular-junction model (Eqs.~\eqref{eqfull}-\eqref{eqHsp}),
\begin{align}
    H & = \frac{\bm{P}^2}{2\bm{m}} + H_s + H_b + H_{sb} + H_{p} + H_{sp}, \label{eqfull}\\
    H_s & = \sum_{j,k} h_{jk} (\bm{R}) c_{j}^{\dagger}c_{k}, \quad H_b = \sum_{g} \sum_{l=L,R}\epsilon_{gl} d_{gl}^{\dagger}d_{gl}, \label{eqhs}
\end{align}
where $c_{j}$ and $d_{gl}$ ($c_{j}^{\dagger}$, and $d_{gl}^{\dagger}$) are the spinless noninteracting fermionic annihilation (creation) operators for the molecule and the electrode modes, respectively.~\cite{smorka24,smorka25} The electrode-molecule exchange is bilinear in nature. The photonic cavity Hamiltonian, $H_p$, is described by a bosonic mode ($a_{\mathrm{cav}}$) and $H_{sp}$ is its coupling with the system degrees of freedom, 
\begin{align}
    H_{p} & = \hbar\omega_{\mathrm{cav}} \left(a^{\dagger}_{\mathrm{cav}}a_{\mathrm{cav}}+ \frac{1}{2}\right), \\
    H_{sp} & = \sum_{j>k} \lambda_{jk}(\bm{R}) \big(c_{j}^{\dagger}c_{k} + c_{k}^{\dagger}c_{j}\big)\big(a^{\dagger}_{\mathrm{cav}} + a_{\mathrm{cav}}\big). \label{eqHsp}
\end{align}
The light-matter coupling, $\lambda$, is, in general, dependent on the nuclear coordinates, $\lambda (\bm{R})$. Following Ref.~\citenum{gu20b}, we have neglected the dipole self-energy term in Eq.~\eqref{eqHsp}, a suitable approximation given that our focus is on population transfer within the strongly driven system.~\cite{rokaj18} 

\begin{figure*}
    \centering
    \includegraphics[width=\textwidth]{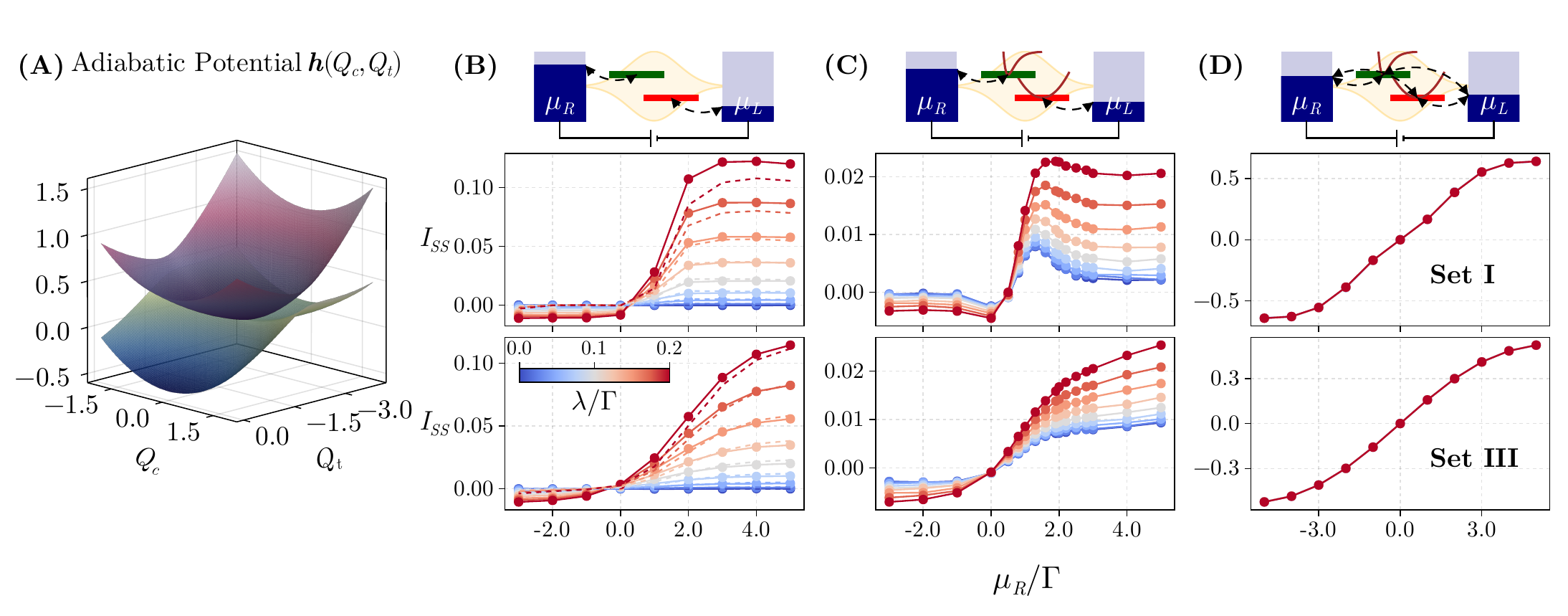} 
    \caption{(A) Adiabatic 2D potential (\textit{c.f.} Eqs.~\eqref{eqPotFull}-\eqref{eqPoth0}) landscape governing nuclear motion ($\bm{R}\equiv\{Q_c,Q_t\}$) in units of coupling strength ($\Gamma$). (B)-(D) Illustrate the three scenarios of the model and two model parameters examined. The top panels show a molecule between left ($\mu_R$) and right ($\mu_L$) electrodes, mediated by a photonic cavity (light yellow). (B) and (C) represent a model with selective connectivity between electrode and molecule, with $\mu_R$ connecting only to the green level and $\mu_L$ connecting only to the red level. In contrast, (D) shows that both electronic levels are connected to two electrodes. (B) lacks a nuclear structure, while both (C) and (D) incorporate a nuclear potential, as illustrated in (A). Same color-scheme has been used in the bottom two rows of panels (B)-(D). The dashed lines in panel (B) display the benchmark HEOM results. The system parameters for Set I and Set III are provided in Table~\ref{tabParams}. Unless otherwise mentioned, we will use the connectivity in panel (C) throughout. Currents are in the units of $e\hbar/\Gamma$.}
    \label{figSteadyStateLambda1}
\end{figure*}

\begin{table}
\caption{Parameter sets used for calculation, all of them are presented in the units of $\Gamma$, the coupling strength between the electrodes and the system. We use $\epsilon_2=-\epsilon_1=\Gamma$, $\Delta \equiv |\epsilon_1-\epsilon_2|$, $\mu_R=-\mu_L$, and $W=10\Gamma$ in all cases, and $\beta$ is the inverse thermal energy. For all the plots, each row corresponds to a specific parameter set.}
\centering
\setlength{\tabcolsep}{6mm}
\renewcommand{\arraystretch}{1.3}
\begin{tabular}{l|c|c}
    \toprule
    & $\beta \Gamma$ &  $\hbar\omega_{\mathrm{cav}}/\Gamma$ \\ \hline\hline
    Set I & 5 & 2 \\ \hline
    Set II & 1 &  2/3 \\ \hline
    Set III & 1 & 2 \\ \botrule
\end{tabular} 
\label{tabParams}
\end{table}

A Lorentzian spectral density was employed,~\cite{haertle13,schinabeck18,batge21} under wide-band approximation.~\cite{verzijl13} The spectral density is expressed as
\begin{equation}
    J_{\mathrm{fermion}}(\epsilon_j) = \frac{\Gamma W^2}{(\epsilon_j-\mu_{L/R})^2 + W^2},\label{eqSpecDen}
\end{equation}
where $\Gamma$ is the coupling strength between the electrode modes, which we assumed to be independent of bias voltage,~\cite{liu17} with chemical potential $\mu_R$ (or, $\mu_L$) and $W$ is the bandwidth.


The time-dependent current can be computed by keeping track of the change in population of the fermionic reservoirs.~\cite{naskar23} We are using a two-dimensional oscillator ($\bm{R}\equiv \{Q_c,Q_t\}$) for the nuclear potential.~\cite{chen16,gu20a,gu20b,cho22}  Numerical details and the form of the potential are provided in \supp{}. For these current parameter choices, there is a conical intersection (CI) at $(Q_t, Q_c)\approx(-2,0)$. The form of $\lambda (\bm{R})$ can be obtained by computing the electrostatic potential across a biased molecular junction using Maxwell's equation or Thomas–Fermi-type screening model.~\cite{nitzan02,pleutin03,shen25}

\begin{figure}[b]
    \centering
    \includegraphics[width=\linewidth]{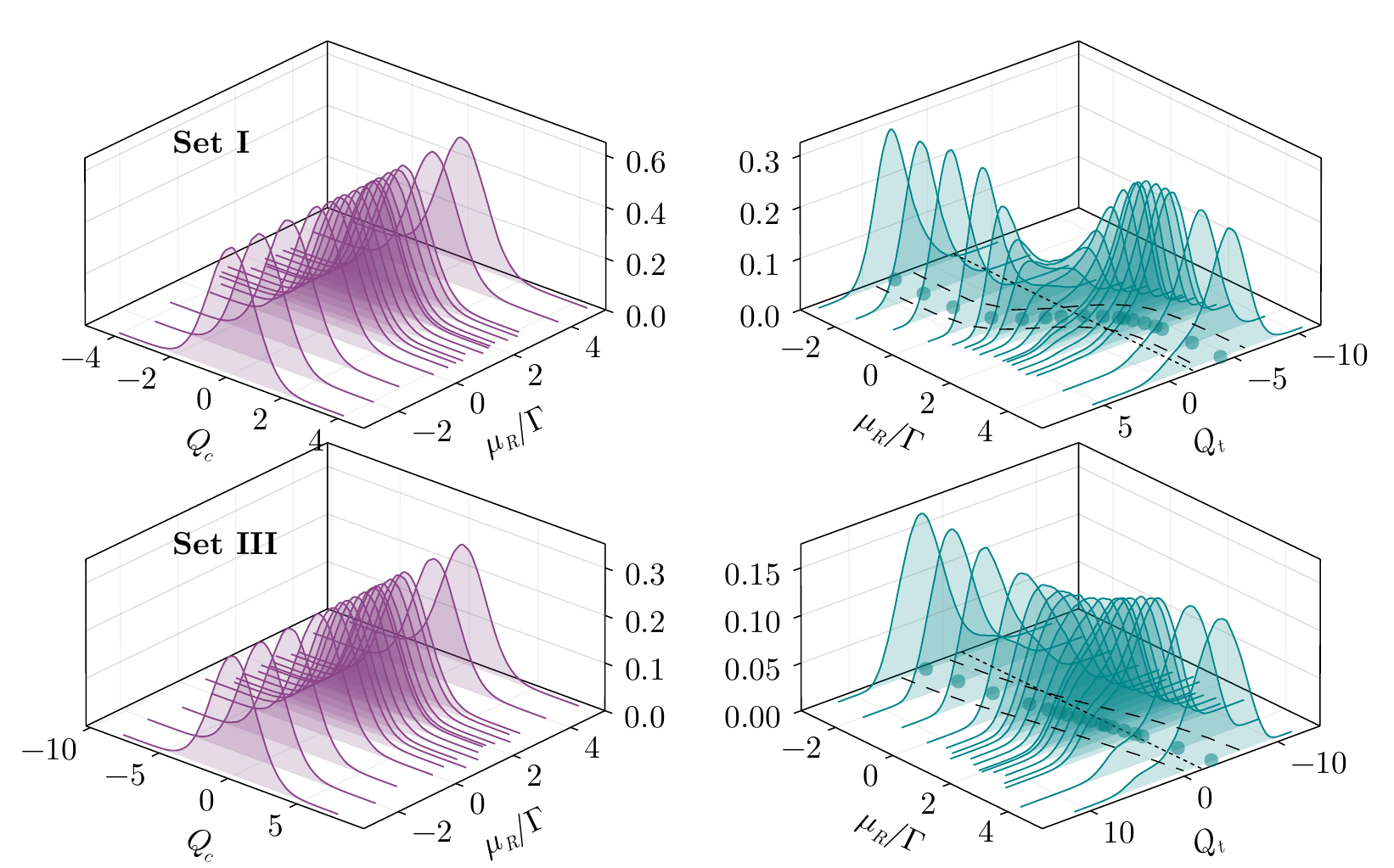}
    \caption{The distributions of the nuclear modes at steady state for Set I \& III, with $\lambda=0.1\Gamma$, for different bias voltages. The left side of the figure shows the $Q_c$ distributions, while the right side displays the $Q_t$ distributions. For the $Q_t$ distributions, the mean and standard deviation are indicated on the base plane by the dotted line and long-dashed boundaries, respectively.}
    \label{figQcQtDist}
\end{figure}


Building on this theoretical framework, we first examine the steady-state current characteristics as a function of light-matter coupling strength ($\lambda$, blue to red) and applied bias, as shown in Figs.~\ref{figSteadyStateLambda1}~(B–D). The three scenarios are motivated by the molecular-junction geometry and the corresponding connectivity to the electrodes, while Sets I and III represent two different temperatures. It is important to note that, in contrast to optically driven plasmonic cavities, excitation in an STM-BJ is generated through resonant transport under applied bias and is interfacial in character, involving hybridized molecule–electrode states. This distinct mechanism motivates the selective electrode–molecule connectivity used in Figs.~\ref{figSteadyStateLambda1}~(B,C), where transport is forced through the light-matter-coupled molecular states. Moreover, because the molecule is mechanically constrained between the two electrodes in an STM-BJ, nuclear motion is often expected to be comparatively restricted, motivating the nuclear-mode-free limit considered in Fig.~\ref{figSteadyStateLambda1}~(B).

In the absence of nuclear modes and in the limit of weak diabatic coupling, as shown in panel (B), light-matter coupling provides the only pathway for population transfer. Therefore, increasing $\lambda$ enhances the coherent transfer between the electronic levels. At large positive bias, the current approaches a plateau, while at negative bias the current remains small as the electronic energy levels are off-resonance with the electrode potentials. In this scenario, the clear separation between the blue (no coupling) and red (maximum coupling) curves provides a clear current signature of strong light-matter coupling.

Introducing the nuclear modes (panel (A)) leads to more complex dynamics in the steady-state current (panel (C)), as the nuclear degrees of freedom introduce additional pathways for population transfer. More precisely, as the bias voltage ($\mu_R$) increases, the steady-state current rises rapidly, followed by either a shallow increase at low $\beta$ (Set III) or a slow decrease at high $\beta$ (Set I). The inflection point also shifts to higher bias with increasing light-matter coupling strength, reflecting the larger adiabatic gap at larger $\lambda$.

The origin of this more complex structure becomes clearer from the steady-state nuclear distributions in Fig.~\ref{figQcQtDist}. When the nuclear distribution samples the CI, nonadiabatic population transfer is enhanced. The coupling mode ($Q_c$) changes only weakly, while the tuning-mode distribution ($Q_t$) changes substantially with bias and becomes narrower as the bias increases. At low temperatures (Set I) the mean of the $Q_t$ distribution is closest to the CI at $Q_t\approx -2$ around the peak of the steady-state current, which facilitates an efficient nonadiabatic population transfer. At larger bias voltages, the distribution shifts away from CI and results in a decreasing current. For Set III, the distribution is broader and remains near the crossing region over a wider range of bias, so the sharp maximum is washed out and the current flattens or continues to increase only slowly. The width of the $Q_c$ distribution provides an additional contribution as broader sampling of $Q_c$ promotes transfer between the two electronic levels. Together, this shows how nuclear motion reshapes the current signature of light-matter coupling.

Finally, Fig.~\ref{figSteadyStateLambda1}~(D) considers nonselective electrode–molecule connectivity, where both electronic levels are coupled to both electrodes. In this case, the $\lambda$-dependent contribution to the steady-state current becomes negligible. Direct electrode–molecule exchange bypasses the polaritonic bridge; this shift reflects a change in connectivity that opens direct transport pathways through the junction, rather than an increase in dissipation.

Hence, these results establish that strong light-matter coupling can leave a fingerprint in the steady-state current, while nuclear motion and electrode connectivity reshape its visibility. The current therefore contains information not only about the cavity coupling and nuclear geometry, but also about the microscopic transport pathways through the junction.

\begin{figure}
    \centering
    \includegraphics[width=\linewidth]{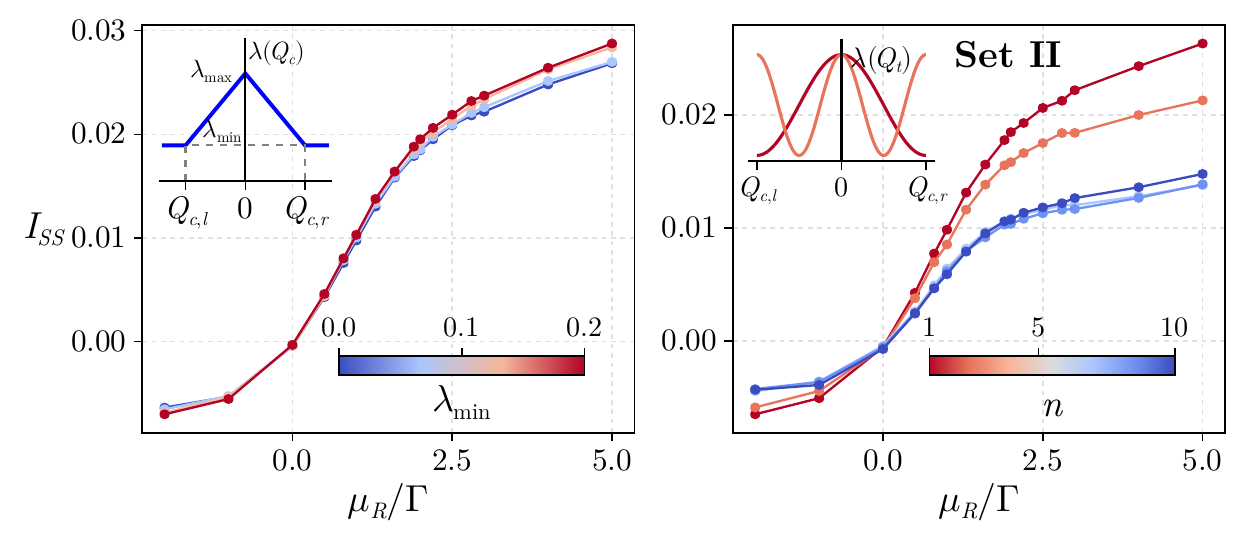}
    \caption{The spatial dependence of $\lambda$ and its impact on current. On the left, the light-matter coupling is a function of the coupling mode, and its shape of $\lambda (Q_c)$ is shown in the inset. We set $\lambda_{\mathrm{max}}=0.2\Gamma$. On the right, it is dependent on the tuning mode. The dependence of $\lambda$ on $Q_t$, for the figure on the right is sinusoidal as shown in the inset, and the exact form is given in the main text. $n$ is the mode number for the sinusoidal form of the light-matter coupling.}
    \label{figCurrentQcQtDependence}
\end{figure}

\begin{figure*}
    \centering
    \includegraphics[width=\linewidth]{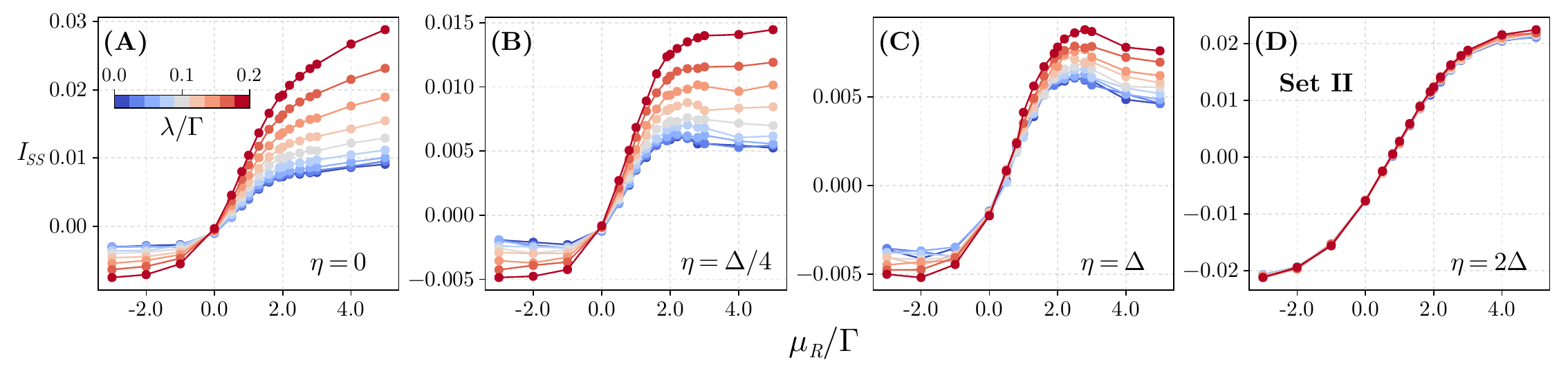}
    \caption{Steady-state currents for varying bias voltage and light-matter coupling in the presence of nuclear modes, similar to Fig.~\ref{figSteadyStateLambda1}, are plotted above for Set II parameters in the presence of a bath. (A) illustrates the variation for a system devoid of bath modes, akin to panel (C) in Fig.~\ref{figSteadyStateLambda1}. (B) presents a similar distribution, but in the presence of nuclear modes where the reorganization energy is in the inverted Marcus regime. (C) exhibits the effect of coupling between the system and the bath modes at activationless crossing, whereas (D) shows the normal Marcus regime. The same colormap is used across all panels. $\Delta$ is the diabatic energy gap of the spinless fermionic system.}
    \label{figExternalBath}
\end{figure*}

Having established this current signature, we next examine how it is affected by the spatial structure of the electromagnetic field (Fig.~\ref{figCurrentQcQtDependence}). This is particularly important in STM-BJs, where the molecule spans the nanocavity formed by the two metallic tips and therefore samples a spatially varying cavity field. Accordingly, the light-matter coupling cannot generally be assumed to be uniform across the molecule.

We consider two distinct scenarios in which the light-matter coupling depends on the nuclear coordinates: coupling-mode dependence, $\lambda(Q_c)$, with a triangular spatial profile (Fig.~\ref{figCurrentQcQtDependence}, left inset), and tuning-mode dependence, $\lambda(Q_t)$, with a sinusoidal spatial profile  (Fig.~\ref{figCurrentQcQtDependence}, right inset). We use the Set II parameters.

When considering the first scenario, \textit{i.e.}, the triangular $\lambda(Q_c)$ profile, the coupling is maximal at $Q_c=0$, and approaches a plateau value $\lambda_{\mathrm{min}}$ away from the center. Increasing $\lambda_{\mathrm{min}}$ leads to a gradual increase in the steady-state current. This trend arises because increasing $\lambda_{\mathrm{min}}$ increases the effective light-matter coupling sampled by the nuclear distribution. Since $Q_c$ remains concentrated around $Q_c=0$ (Fig.~\ref{figQcQtDist}), where the coupling is strongest, the overall steady-state current increases.

This behavior changes qualitatively when we consider the second scenario, \textit{i.e.}, the sinusoidal profile $\lambda (Q_t) = 0.1 \left(\cos \left(\frac{Q_t}{Q_{t,\mathrm{max}}}n\pi\right)+1 \right)$ is the maximum lateral extent of the tuning mode. Here, we explore how varying the integer $n$ changes the steady-state current and find that the current does not vary monotonically with $n$. Instead, because the tuning-mode distribution is more localized and shifts significantly with bias voltage, the current becomes sensitive to the local structure of $\lambda(Q_t)$.

Furthermore, increasing $n$ initially results in a lower steady-state current, as it introduces additional nodes in $\lambda(Q_t)$ and effectively lowers the average coupling strength. Because the nuclear distribution is localized, the current becomes sensitive to the local gradient of the coupling field rather than its global average. Consequently, $Q_t$ frequently lands in a node region of $\lambda(Q_t)$ for a given $n$, reversing the ordering of the current curves such that a more structured coupling profile yields a smaller steady-state current.

Hence, the spatial structure of the electromagnetic field can qualitatively reshape the transport signature of strong light-matter coupling. Interestingly, the steady-state current therefore does not only report the strength of the cavity-molecule coupling, but also conveys its spatial profile.


Finally, we explore the role of the solvent environment on the steady-state current. This choice is motivated by the fact that STM-BJ measurements are commonly performed in solution, where solvent interactions are known to influence charge transport.~\cite{Kirchberg20,chen23} We investigate the transport characteristics in the presence of an external bosonic bath. The external bath is taken to be a collection of oscillators linearly coupled to the system modes,~\cite{segal10,sowa18}
\begin{align}
    H_{\mathrm{sol}} & = \sum_{\alpha=1} \hbar \omega_{\mathrm{sol},\alpha} \bigg( a_{\mathrm{sol},\alpha}^{\dagger}a_{\mathrm{sol},\alpha} + \frac{1}{2} \bigg), \\
    H_{s,\mathrm{sol}} & = \sum_{\alpha=1} \frac{g_{\alpha}}{\sqrt{2}}(a_{\mathrm{sol},\alpha}^{\dagger}+a_{\mathrm{sol},\alpha})(c_1^{\dagger}c_1-c_2^{\dagger}c_2),
\end{align}
where  $\omega_{\mathrm{sol},\alpha}$ are the frequencies of the solvent modes and $g_{\alpha}$ are the coupling constants to the bath modes with the fermionic system levels. The bosonic modes are sampled from a Lorentzian spectral density, $J_{\mathrm{boson}} (\omega) = \frac{2\eta \omega_c \omega}{\omega^2+\omega_c^2}$, where $\eta$ is the bath reorganization energy, and $\omega_c$ is the peak frequency of the spectral distribution. A discretization scheme for the bath modes is described in the \supp{}, Sec.~\ref{appInitialCond}.

We consider four solvent-coupling regimes~\cite{Migliore11} in Fig.~\ref{figExternalBath}: the solvent-free limit in panel (A), the inverted Marcus regime in panel (B), the activationless regime in panel (C), and the normal Marcus regime~\cite{marcus85} in panel (D). We use the Set II parameters. The presence of the bath generally reduces the overall transport rate, with this suppression becoming more pronounced in the inverted Marcus regime as the system–bath coupling increases, as shown in panel (B). Near the activationless condition in panel (C), the current develops a pronounced peak and the current–voltage response becomes increasingly symmetric with respect to bias. This evolution suggests that the solvent bath does not simply attenuate transport, but also reshapes the available transfer pathways. In the normal Marcus regime, panel (D), strong environmental coupling opens an additional transfer pathway, causing the current to become largely insensitive to the light-matter coupling and to resemble the cavity-free response in Fig.~\ref{figSteadyStateLambda1}~(D). Hence, solvent interactions can substantially reshape or even suppress the current signature of strong light-matter coupling, further highlighting the sensitivity of steady-state current to the local molecular environment.


In conclusion, this work establishes steady-state current as a sensitive indicator of strong light-matter coupling in molecular junctions, complementing optical spectroscopy. Recent STM-BJ experiments have shown that electroluminescence spectra can display the characteristic spectroscopic signature of strong light-matter coupling, namely Rabi splitting.~\cite{paoletta24,chikkaraddy16} Because STM-BJs also provide direct access to the junction current, we ask whether the resulting polaritonic states can leave a measurable fingerprint in the steady-state current. Interestingly, we find that they can, opening a complementary route to probing strong light-matter coupling beyond optical spectra.

We evaluate the robustness of this current signature across different junction environments. We first consider selective electrode–molecule connectivity and restricted nuclear motion, which are both well motivated for voltage-driven STM-BJs. In this limit, the light-matter-coupling-dependent current signature is particularly clear. We then introduce nuclear motion, spatially structured electromagnetic fields, and solvent coupling. We demonstrate that strong light-matter coupling  produces characteristic signatures in the current that, while subject to reshaping by the microscopic environment, remain fundamentally sensitive to the coupled state.~\cite{mondal24,sandik25,aradhya13} Rather than viewing molecular vibrations as added complexity, our results show that they can provide additional microscopic information about the coupled molecular system. This opens a route toward using electrical transport not only to detect strong light-matter coupling, but also to probe and ultimately control polaritonic dynamics at the single-molecule level. Future research will expand this framework to incorporate complexities such as an accurate representation of the nuclear potential, electron-phonon interactions, spin-orbit coupling,~\cite{fay21,li23,fay23} and spin dynamics, and electron-electron correlations, thereby enabling the design of complex, multifunctional nanoscale devices.

\vspace{0.1in}
\noindent
\textbf{Data Availability:} Codes for reproducing the figures in this manuscript are available at \href{https://github.com/kritanjan-polley/stm-bj-steady-state}{Github repository}.

\vspace{0.1in}
\noindent
\textbf{\supp{}:} The details of the semiclassical approaches, propagation details, comparisons with HEOM and master equation approach, and gauge transformed Hamiltonians are provided in the \supp{}.

\vspace{0.1in}
\noindent
\textbf{Acknowledgments:} This work was supported by a grant from the Simons Foundation [MPS-T-MPS-00839534, MET] and the U.S. DOE, Office of Science, BES (Early Career Award No. DE-SC0026328), as managed by the CPIMS program.

\bibliography{reference}

@article{sowa18,
    author = {Sowa, Jakub K. and Mol, Jan A. and Briggs, G. Andrew D. and Gauger, Erik M.},
    title = {Beyond Marcus theory and the Landauer-Büttiker approach in molecular junctions: A unified framework},
    journal = {J. Chem. Phys.},
    volume = {149},
    number = {15},
    pages = {154112},
    year = {2018},
    month = {10},
    issn = {0021-9606},
    doi = {10.1063/1.5049537},
    url = {https://doi.org/10.1063/1.5049537},
}

@article{Migliore11,
    author = {Migliore, Agostino and Nitzan, Abraham},
    title = {Nonlinear Charge Transport in Redox Molecular Junctions: A Marcus Perspective},
    journal = {ACS Nano},
    volume = {5},
    number = {8},
    pages = {6669-6685},
    year = {2011},
    month = {08},
    issn = {1936-0851},
    doi = {10.1021/nn202206e},
    url = {https://doi.org/10.1021/nn202206e},
}

@article{Kirchberg20,
    author = {Kirchberg, Henning and Thorwart, Michael and Nitzan, Abraham},
    title = {Charge Transfer through Redox Molecular Junctions in Nonequilibrated Solvents},
    journal = {J. Phys. Chem. Lett.},
    volume = {11},
    number = {5},
    pages = {1729-1737},
    year = {2020},
    month = {03},
    issn = {1948-7185},
    doi = {10.1021/acs.jpclett.0c00118},
    url = {https://doi.org/10.1021/acs.jpclett.0c00118},
}

@article{chen23,
  title={Reactions in single-molecule junctions},
  author={Chen, Hongliang and Jia, Chuancheng and Zhu, Xin and Yang, Chen and Guo, Xuefeng and Stoddart, J Fraser},
  journal={Nat. Rev. Mater.},
  volume={8},
  number={3},
  pages={165--185},
  year={2023},
  doi={10.1038/s41578-022-00506-0},
  publisher={Nature Publishing Group UK London}
}

@article{tanimura20,
    author = {Tanimura, Yoshitaka},
    title = {Numerically “exact” approach to open quantum dynamics: The hierarchical equations of motion (HEOM)},
    journal = {J. Chem. Physics.},
    volume = {153},
    number = {2},
    pages = {020901},
    year = {2020},
    month = {07},
    issn = {0021-9606},
    doi = {10.1063/5.0011599},
    url = {https://doi.org/10.1063/5.0011599},
}

@misc{wang26,
      title={Decoherence Tuning in Voltage-Driven Plasmonic Cavities}, 
      author={Yuchen Wang and Oliver Tan and Norah M. Hoffmann},
      year={2026},
      eprint={2609.28287},
      archivePrefix={arXiv},
      primaryClass={quant-ph},
      url={https://arxiv.org/abs/2609.28287}, 
}

@article{cui22,
    author = {Cui, Bingyu and Nizan, Abraham},
    title = {Collective response in light–matter interactions: The interplay between strong coupling and local dynamics},
    journal = {J. Chem. Phys.},
    volume = {157},
    number = {11},
    pages = {114108},
    year = {2022},
    month = {09},
    issn = {0021-9606},
    doi = {10.1063/5.0101528},
    url = {https://doi.org/10.1063/5.0101528},
}

@article{Meisner12,
    author = {Meisner, Jeffrey S. and Ahn, Seokhoon and Aradhya, Sriharsha V. and Krikorian, Markrete and Parameswaran, Radha and Steigerwald, Michael and Venkataraman, Latha and Nuckolls, Colin},
    title = {Importance of Direct Metal-pi Coupling in Electronic Transport Through Conjugated Single-Molecule Junctions},
    journal = {J. Am. Chem. Soc.},
    volume = {134},
    number = {50},
    pages = {20440-20445},
    year = {2012},
    month = {12},
    issn = {0002-7863},
    doi = {10.1021/ja308626m},
    url = {https://doi.org/10.1021/ja308626m},
}

@article{Galbiati12,
  title = {Polariton Condensation in Photonic Molecules},
  author = {Galbiati, Marta and Ferrier, Lydie and Solnyshkov, Dmitry D. and Tanese, Dimitrii and Wertz, Esther and Amo, Alberto and Abbarchi, Marco and Senellart, Pascale and Sagnes, Isabelle and Lema\^{\i}tre, Aristide and Galopin, Elisabeth and Malpuech, Guillaume and Bloch, Jacqueline},
  journal = {Phys. Rev. Lett.},
  volume = {108},
  issue = {12},
  pages = {126403},
  numpages = {5},
  year = {2012},
  month = {Mar},
  publisher = {American Physical Society},
  doi = {10.1103/PhysRevLett.108.126403},
  url = {https://link.aps.org/doi/10.1103/PhysRevLett.108.126403}
}

@article{low17,
  title={Polaritons in layered two-dimensional materials},
  author={Low, Tony and Chaves, Andrey and Caldwell, Joshua D and Kumar, Anshuman and Fang, Nicholas X and Avouris, Phaedon and Heinz, Tony F and Guinea, Francisco and Martin-Moreno, Luis and Koppens, Frank},
  journal={Nature Mater.},
  volume={16},
  number={2},
  pages={182--194},
  year={2017},
  doi={10.1038/nmat4792},
  publisher={Nature Publishing Group UK London}
}

@article{thomas16,
author = {Thomas, Anoop and George, Jino and Shalabney, Atef and Dryzhakov, Marian and Varma, Sreejith J. and Moran, Joseph and Chervy, Thibault and Zhong, Xiaolan and Devaux, Eloïse and Genet, Cyriaque and Hutchison, James A. and Ebbesen, Thomas W.},
title = {Ground-State Chemical Reactivity under Vibrational Coupling to the Vacuum Electromagnetic Field},
journal = {Angew. Chem. Int. Ed.},
volume = {55},
number = {38},
pages = {11462-11466},
doi = {https://doi.org/10.1002/anie.201605504},
year = {2016}
}

@article{xiang24,
    author = {Xiang, Bo and Xiong, Wei},
    title = {Molecular Polaritons for Chemistry, Photonics and Quantum Technologies},
    journal = {Chem. Rev.},
    volume = {124},
    number = {5},
    pages = {2512-2552},
    year = {2024},
    month = {03},
    issn = {0009-2665},
    doi = {10.1021/acs.chemrev.3c00662},
    url = {https://doi.org/10.1021/acs.chemrev.3c00662},
}

@article{hu24,
  title={Robust consistent single quantum dot strong coupling in plasmonic nanocavities},
  author={Hu, Shu and Huang, Junyang and Arul, Rakesh and S{\'a}nchez-Iglesias, Ana and Xiong, Yuling and Liz-Marz{\'a}n, Luis M and Baumberg, Jeremy J},
  journal={Nat. Commun.},
  volume={15},
  number={1},
  pages={6835},
  year={2024},
  doi={10.1038/s41467-024-51170-7},
  publisher={Nature Publishing Group UK London}
}

@article{wang22,
  title={Plasmonic phenomena in molecular junctions: principles and applications},
  author={Wang, Maoning and Wang, Tao and Ojambati, Oluwafemi S and Duffin, Thorin Jake and Kang, Keehoon and Lee, Takhee and Scheer, Elke and Xiang, Dong and Nijhuis, Christian A},
  journal={Nat. Rev. Chem.},
  volume={6},
  number={10},
  pages={681--704},
  doi={10.1038/s41570-022-00423-4},
  year={2022},
  publisher={Nature Publishing Group UK London}
}

@article{baumberg19,
  title={Extreme nanophotonics from ultrathin metallic gaps},
  author={Baumberg, Jeremy J and Aizpurua, Javier and Mikkelsen, Maiken H and Smith, David R},
  journal={Nat. Mater.},
  volume={18},
  number={7},
  pages={668--678},
  year={2019},
  doi={10.1038/s41563-019-0290-y},
  publisher={Nature Publishing Group UK London}
}

@article{polley20,
    author = {Polley, Kritanjan and Loring, Roger F.},
    title = {Spectroscopic response theory with classical mapping Hamiltonians},
    journal = {J. Chem. Phys.},
    volume = {153},
    number = {20},
    pages = {204103},
    year = {2020},
    month = {11},
    issn = {0021-9606},
    url = {https://doi.org/10.1063/5.0029231},
}

@article{jung23,
    author = {Jung, Kenneth A. and Kelly, Joseph and Markland, Thomas E.},
    title = "{Electron transfer at electrode interfaces via a straightforward quasiclassical fermionic mapping approach}",
    journal = {J. Chem. Phys.},
    volume = {159},
    number = {1},
    pages = {014109},
    year = {2023},
    month = {07},
    doi = {10.1063/5.0156136},
}

@article{meyer79a,
  title={A classical analog for electronic degrees of freedom in nonadiabatic collision processes},
  author={Meyer, Hans-Dieter and Miller, William H},
  journal={J. Chem. Phys.},
  volume={70},
  number={7},
  pages={3214--3223},
  year={1979},
  publisher={American Institute of Physics},
  url={https://doi.org/10.1063/1.437910}
}

@article{meyer79b,
  title={Classical models for electronic degrees of freedom: Derivation via spin analogy and application to F*+ H2→ F+ H2},
  author={Meyer, Hans-Dieter and Miller, William H},
  journal={J. Chem. Phys.},
  volume={71},
  number={5},
  pages={2156--2169},
  year={1979},
  publisher={American Institute of Physics}
}

@article{stock97,
  title={Semiclassical description of nonadiabatic quantum dynamics},
  author={Stock, Gerhard and Thoss, Michael},
  journal={Phys. Rev. Lett.},
  volume={78},
  number={4},
  pages={578},
  year={1997},
  publisher={APS}
}

@article{thoss99,
  title={Mapping approach to the semiclassical description of nonadiabatic quantum dynamics},
  author={Thoss, Michael and Stock, Gerhard},
  journal={Phys. Rev. A},
  volume={59},
  number={1},
  pages={64},
  year={1999},
  publisher={APS}
}

@article{levy19,
  title={A complete quasiclassical map for the dynamics of interacting fermions},
  author={Levy, Amikam and Dou, Wenjie and Rabani, Eran and Limmer, David T},
  journal={J. Chem. Phys.},
  volume={150},
  number={23},
  year={2019},
  pages={234112},
  publisher={AIP Publishing},
  url={https://pubs.aip.org/aip/jcp/article/150/23/234112/197664}
}

@article{li12,
  title={A Cartesian classical second-quantized many-electron Hamiltonian, for use with the semiclassical initial value representation},
  author={Li, Bin and Miller, William H},
  journal={J. Chem. Phys.},
  volume={137},
  number={15},
  year={2012},
  pages={154107},
  publisher={AIP Publishing},
  url={https://pubs.aip.org/aip/jcp/article/137/15/154107/976368}
}

@article{li14,
  title={A quasi-classical mapping approach to vibrationally coupled electron transport in molecular junctions},
  author={Li, Bin and Wilner, Eli Y and Thoss, Michael and Rabani, Eran and Miller, William H},
  journal={J. Chem. Phys.},
  volume={140},
  number={10},
  year={2014},
  pages={104110},
  publisher={AIP Publishing},
  url={https://pubs.aip.org/aip/jcp/article/140/10/104110/351360}
}

@article{li13,
  title={A Cartesian quasi-classical model to nonequilibrium quantum transport: The Anderson impurity model},
  author={Li, Bin and Levy, Tal J and Swenson, David WH and Rabani, Eran and Miller, William H},
  journal={J. Chem. Phys.},
  volume={138},
  number={10},
  year={2013},
  pages={104110},
  publisher={AIP Publishing},
  url={https://pubs.aip.org/aip/jcp/article/138/10/104110/192822}
}

@article{huang23,
  doi = {10.1038/s42005-023-01427-2},
  url = {https://doi.org/10.1038/s42005-023-01427-2},
  year = {2023},
  month = {Oct},
  publisher = {Nature Portfolio},
  volume = {6},
  number = {1},
  pages = {313},
  author = {Huang, Yi-Te and Kuo, Po-Chen and Lambert, Neill and Cirio, Mauro and Cross, Simon and Yang, Shen-Liang and Nori, Franco and Chen, Yueh-Nan},
  title = {An efficient {J}ulia framework for hierarchical equations of motion in open quantum systems},
  journal = {Commun. Phys.}
}

@article{smorka25,
  title={Influence of nonequilibrium vibrational dynamics on spin selectivity in chiral molecular junctions},
  author={Smorka, Rudolf and Rudge, Samuel L and Thoss, Michael},
  journal={J. Chem. Phys.},
  volume={162},
  number={1},
  year={2025},
  publisher={AIP Publishing},
  url={https://doi.org/10.1063/5.0235411}
}

@article{naskar23,
  title={Chiral-induced spin selectivity and non-equilibrium spin accumulation in molecules and interfaces: A first-principles study},
  author={Naskar, Sumit and Mujica, Vladimiro and Herrmann, Carmen},
  journal={J. Phys. Chem. Lett.},
  volume={14},
  number={3},
  pages={694--701},
  year={2023},
  publisher={ACS Publications},
  url={https://doi.org/10.1021/acs.jpclett.2c03747}
}

@article{smorka24,
  title={Dynamics of spin relaxation in nonequilibrium magnetic nanojunctions},
  author={Smorka, Rudolf and Thoss, Michael and {\v{Z}}onda, Martin},
  journal={New J. Phys.},
  volume={26},
  number={1},
  pages={013056},
  year={2024},
  publisher={IOP Publishing},
  url={https://10.1088/1367-2630/ad1fa9}
}

@article{runeson19,
    author = {Runeson, Johan E. and Richardson, Jeremy O.},
    title = {Spin-mapping approach for nonadiabatic molecular dynamics},
    journal = {J. Chem. Phys.},
    volume = {151},
    number = {4},
    pages = {044119},
    year = {2019},
    month = {07},
    issn = {0021-9606},
    doi = {10.1063/1.5100506},
    url = {https://doi.org/10.1063/1.5100506},
}

@article{langmann15,
  title={Construction by bosonization of a fermion-phonon model},
  author={Langmann, Edwin and Moosavi, Per},
  journal={J. Math. Phys.},
  volume = {56},
  number = {9},
  pages = {091902},
  year = {2015},
  month = {09},
  url = {https://doi.org/10.1063/1.4930299},
  publisher={AIP Publishing}
}

@article{sun21,
    author = {Sun, Jing and Sasmal, Sudip and Vendrell, Oriol},
    title = {A bosonic perspective on the classical mapping of fermionic quantum dynamics},
    journal = {J. Chem. Phys.},
    volume = {155},
    number = {13},
    pages = {134110},
    year = {2021},
    month = {10},
    issn = {0021-9606},
    doi = {10.1063/5.0066740},
    url = {https://doi.org/10.1063/5.0066740},
}

@article{montoya18,
  title={On the exact continuous mapping of fermions},
  author={Montoya-Castillo, Andr{\'e}s and Markland, Thomas E},
  journal={Sci. Rep.},
  volume={8},
  number={1},
  pages={12929},
  year={2018},
  publisher={Nature Publishing Group UK London},
  doi={https://doi.org/10.1038/s41598-018-31162-6}
}

@article{montoya23,
  title={A derivation of the conditions under which bosonic operators exactly capture fermionic structure and dynamics},
  author={Montoya-Castillo, Andr{\'e}s and Markland, Thomas E},
  journal={J. Chem. Phys.},
  volume={158},
  number={9},
  year={2023},
  url={https://doi.org/10.1063/5.0138664},
  publisher={AIP Publishing}
}

@article{gu20a,
  title={Manipulating nonadiabatic conical intersection dynamics by optical cavities},
  author={Gu, Bing and Mukamel, Shaul},
  journal={Chem. Sci.},
  volume={11},
  number={5},
  pages={1290--1298},
  year={2020},
  doi={10.1039/C9SC04992D},
  publisher={Royal Society of Chemistry}
}

@article{gu20b,
  title={Cooperative conical intersection dynamics of two pyrazine molecules in an optical cavity},
  author={Gu, Bing and Mukamel, Shaul},
  journal={J. Phys. Chem. Lett.},
  volume={11},
  number={14},
  pages={5555--5562},
  year={2020},
  doi={10.1021/acs.jpclett.0c00381},
  publisher={ACS Publications}
}

@article{cho22,
  title={{Optical cavity manipulation and nonlinear UV molecular spectroscopy of conical intersections in pyrazine}},
  author={Cho, Daeheum and Gu, Bing and Mukamel, Shaul},
  journal={J. Am. Chem. Soc.},
  volume={144},
  number={17},
  pages={7758--7767},
  year={2022},
  publisher={ACS Publications},
  doi={https://doi.org/10.1021/jacs.2c00921}
}

@article{woolley20,
  title={Power-Zienau-Woolley representations of nonrelativistic QED for atoms and molecules},
  author={Woolley, R Guy},
  journal={Phys. Rev. Research},
  volume={2},
  number={1},
  pages={013206},
  year={2020},
  publisher={APS},
  doi={https://doi.org/10.1103/PhysRevResearch.2.013206}
}

@article{babiker83,
  title={Derivation of the Power-Zienau-Woolley Hamiltonian in quantum electrodynamics by gauge transformation},
  author={Babiker, M and Loudon, Rodney},
  journal={Proc. A},
  volume={385},
  number={1789},
  pages={439--460},
  year={1983},
  publisher={The Royal Society London},
  doi={https://doi.org/10.1098/rspa.1983.0022}
}

@article{batge21,
  title = {Nonequilibrium open quantum systems with multiple bosonic and fermionic environments: A hierarchical equations of motion approach},
  author = {B\"atge, J. and Ke, Y. and Kaspar, C. and Thoss, M.},
  journal = {Phys. Rev. B},
  volume = {103},
  issue = {23},
  pages = {235413},
  numpages = {11},
  year = {2021},
  month = {Jun},
  publisher = {American Physical Society},
  doi = {10.1103/PhysRevB.103.235413},
}

@article{schinabeck18,
  title = {Hierarchical quantum master equation approach to electronic-vibrational coupling in nonequilibrium transport through nanosystems: Reservoir formulation and application to vibrational instabilities},
  author = {Schinabeck, C. and H\"artle, R. and Thoss, M.},
  journal = {Phys. Rev. B},
  volume = {97},
  issue = {23},
  pages = {235429},
  numpages = {21},
  year = {2018},
  month = {Jun},
  publisher = {American Physical Society},
  doi = {10.1103/PhysRevB.97.235429},
}

@article{xiang16,
  title={Molecular-scale electronics: from concept to function},
  author={Xiang, Dong and Wang, Xiaolong and Jia, Chuancheng and Lee, Takhee and Guo, Xuefeng},
  journal={Chem. Rev.},
  volume={116},
  number={7},
  pages={4318--4440},
  year={2016},
  doi = {10.1021/acs.chemrev.5b00680},
  publisher={ACS Publications}
}

@article{su16,
  title={Chemical principles of single-molecule electronics},
  author={Su, Timothy A and Neupane, Madhav and Steigerwald, Michael L and Venkataraman, Latha and Nuckolls, Colin},
  journal={Nat. Rev. Mater.},
  volume={1},
  number={3},
  pages={1--15},
  year={2016},
  url={https://doi.org/10.1038/natrevmats.2016.2},
  publisher={Nature Publishing Group}
}

@article{kamenetska10,
  title={Conductance and geometry of pyridine-linked single-molecule junctions},
  author={Kamenetska, M and Quek, Su Ying and Whalley, AC and Steigerwald, ML and Choi, HJ and Louie, Steven G and Nuckolls, C and Hybertsen, MS and Neaton, JB and Venkataraman, Latha},
  journal={J. Am. Chem. Soc.},
  volume={132},
  number={19},
  pages={6817--6821},
  doi={10.1021/ja1015348},
  year={2010},
  publisher={ACS Publications}
}

@article{naaman19,
  title={Chiral molecules and the electron spin},
  author={Naaman, Ron and Paltiel, Yossi and Waldeck, David H},
  journal={Nat. Rev. Chem.},
  volume={3},
  number={4},
  pages={250--260},
  year={2019},
  publisher={Nature Publishing Group UK London},
  doi={10.1038/s41570-019-0087-1}
}

@article{dief23,
  title={Advances in single-molecule junctions as tools for chemical and biochemical analysis},
  author={Dief, Essam M and Low, Paul J and D{\'\i}ez-P{\'e}rez, Ismael and Darwish, Nadim},
  journal={Nat. Chem.},
  volume={15},
  number={5},
  pages={600--614},
  year={2023},
  publisher={Nature Publishing Group UK London},
  doi={10.1038/s41557-023-01178-1}
}

@article{chen20,
  title = {Exact bosonization in arbitrary dimensions},
  author = {Chen, Yu-An},
  journal = {Phys. Rev. Res.},
  volume = {2},
  issue = {3},
  pages = {033527},
  numpages = {10},
  year = {2020},
  month = {Sep},
  publisher = {American Physical Society},
  doi = {10.1103/PhysRevResearch.2.033527}
}

@article{dhar06,
doi = {10.1088/1126-6708/2006/01/118},
year = {2006},
month = {jan},
publisher = {},
volume = {2006},
number = {01},
pages = {118},
author = {Avinash Dhar and Gautam Mandal and Nemani V. Suryanarayana},
title = {Exact operator bosonization of finite number
of fermions in one space dimension},
journal = {Journal of High Energy Physics}
}

@article{segal10,
  title = {Numerically exact path-integral simulation of nonequilibrium quantum transport and dissipation},
  author = {Segal, Dvira and Millis, Andrew J. and Reichman, David R.},
  journal = {Phys. Rev. B},
  volume = {82},
  issue = {20},
  pages = {205323},
  numpages = {13},
  year = {2010},
  month = {Nov},
  publisher = {American Physical Society},
  doi = {10.1103/PhysRevB.82.205323}
}

@article{paoletta24,
author = {Paoletta, Angela L. and Hoffmann, Norah M. and Cheng, Daniel W. and York, Emma and Xu, Ding and Zhang, Boyuan and Delor, Milan and Berkelbach, Timothy C. and Venkataraman, Latha},
title = {Plasmon-Exciton Strong Coupling in Single-Molecule Junction Electroluminescence},
journal = {J. Am. Chem. Soc.},
volume = {146},
number = {50},
pages = {34394-34400},
year = {2024},
doi = {10.1021/jacs.4c09782}
}

@article{chikkaraddy16,
  title={Single-molecule strong coupling at room temperature in plasmonic nanocavities},
  author={Chikkaraddy, Rohit and De Nijs, Bart and Benz, Felix and Barrow, Steven J and Scherman, Oren A and Rosta, Edina and Demetriadou, Angela and Fox, Peter and Hess, Ortwin and Baumberg, Jeremy J},
  journal={Nature},
  volume={535},
  number={7610},
  pages={127--130},
  doi={10.1038/nature17974},
  year={2016},
  publisher={Nature Publishing Group UK London}
}

@article{benz16,
  title={Single-molecule optomechanics in “picocavities”},
  author={Benz, Felix and Schmidt, Mikolaj K and Dreismann, Alexander and Chikkaraddy, Rohit and Zhang, Yao and Demetriadou, Angela and Carnegie, Cloudy and Ohadi, Hamid and De Nijs, Bart and Esteban, Ruben and others},
  journal={Science},
  volume={354},
  number={6313},
  pages={726--729},
  doi={10.1126/science.aah5243},
  year={2016},
  publisher={American Association for the Advancement of Science}
}

@article{kuisma22,
author = {Kuisma, Mikael and Rousseaux, Benjamin and Czajkowski, Krzysztof M. and Rossi, Tuomas P. and Shegai, Timur and Erhart, Paul and Antosiewicz, Tomasz J.},
title = {Ultrastrong Coupling of a Single Molecule to a Plasmonic Nanocavity: A First-Principles Study},
journal = {ACS Photonics},
volume = {9},
number = {3},
pages = {1065-1077},
year = {2022},
doi = {10.1021/acsphotonics.2c00066},
}

@article{martin20,
  title={Unveiling the radiative local density of optical states of a plasmonic nanocavity by STM},
  author={Mart{\'\i}n-Jim{\'e}nez, Alberto and Fern{\'a}ndez-Dom{\'\i}nguez, Antonio I and Lauwaet, Koen and Granados, Daniel and Miranda, Rodolfo and Garc{\'\i}a-Vidal, Francisco J and Otero, Roberto},
  journal={Nat. Commun.},
  volume={11},
  number={1},
  pages={1021},
  year={2020},
  doi={10.1038/s41467-020-14827-7},
  publisher={Nature Publishing Group UK London}
}

@article{zheng25,
  title={Active control of excitonic strong coupling and electroluminescence in electrically driven plasmonic nanocavities},
  author={Zheng, Junsheng and Krasavin, Alexey V and Yang, Ruoxue and Wang, Zhenxin and Feng, Yuanjia and Tang, Longhua and Li, Linjun and Guo, Xin and Dai, Daoxin and Zayats, Anatoly V and others},
  journal={Sci. Adv.},
  volume={11},
  number={22},
  pages={eadt9808},
  year={2025},
  doi={10.1126/sciadv.adt9808},
  publisher={American Association for the Advancement of Science}
}

@article{bitton22,
  title={Plasmonic cavities and individual quantum emitters in the strong coupling limit},
  author={Bitton, Ora and Haran, Gilad},
  journal={Acc. Chem. Res.},
  volume={55},
  number={12},
  pages={1659--1668},
  year={2022},
  doi={10.1021/acs.accounts.2c00028},
  publisher={ACS Publications}
}

@article{thoss18,
    author = {Thoss, Michael and Evers, Ferdinand},
    title = {Perspective: Theory of quantum transport in molecular junctions},
    journal = {J. Chem. Phys.},
    volume = {148},
    number = {3},
    pages = {030901},
    year = {2018},
    month = {01},
    issn = {0021-9606},
    doi = {10.1063/1.5003306},
}

@article{haertle13,
  title = {Decoherence and lead-induced interdot coupling in nonequilibrium electron transport through interacting quantum dots: A hierarchical quantum master equation approach},
  author = {H\"artle, R. and Cohen, G. and Reichman, D. R. and Millis, A. J.},
  journal = {Phys. Rev. B},
  volume = {88},
  issue = {23},
  pages = {235426},
  numpages = {20},
  year = {2013},
  month = {Dec},
  publisher = {American Physical Society},
  doi = {10.1103/PhysRevB.88.235426},
  url = {https://link.aps.org/doi/10.1103/PhysRevB.88.235426}
}

@article{jin08,
    author = {Jin, Jinshuang and Zheng, Xiao and Yan, YiJing},
    title = {Exact dynamics of dissipative electronic systems and quantum transport: Hierarchical equations of motion approach},
    journal = {J. Chem. Phys.},
    volume = {128},
    number = {23},
    pages = {234703},
    year = {2008},
    month = {06},
    issn = {0021-9606},
    doi = {10.1063/1.2938087},
}

@article{zhao26,
	author = {Zhao, Yuxin and Liang, Wenjie and Zhao, Yanli},
	doi = {10.1038/s42254-025-00888-4},
	isbn = {2522-5820},
	journal = {Nat. Rev. Phys.},
	number = {1},
	pages = {9--26},
	title = {Quantum correlation behaviour in single-molecule junctions},
	volume = {8},
	year = {2026},
}

@article{gao24,
  title={Technologies for investigating single-molecule chemical reactions},
  author={Gao, Chunyan and Gao, Qinghua and Zhao, Cong and Huo, Yani and Zhang, Zhizhuo and Yang, Jinlong and Jia, Chuancheng and Guo, Xuefeng},
  journal={Natl. Sci. Rev.},
  volume={11},
  number={8},
  pages={nwae236},
  year={2024},
  doi={10.1093/nsr/nwae236},
  publisher={Oxford University Press}
}

@article{prana24,
  title={Lewis-acid mediated reactivity in single-molecule junctions},
  author={Prana, Jazmine and Kim, Leopold and Czyszczon-Burton, Thomas M and Homann, Grace and Chen, Sully F and Miao, Zelin and Camarasa-Gomez, Maria and Inkpen, Michael S},
  journal={J. Am. Chem. Soc.},
  volume={146},
  number={48},
  pages={33265--33275},
  year={2024},
  doi={10.1021/jacs.4c14176},
  publisher={ACS Publications}
}

@article{york25,
  title={Tuning Conductance in BODIPY-Based Single-Molecule Junctions},
  author={York, Emma and Stone, Ilana and Shi, Wanzhuo and Roy, Xavier and Venkataraman, Latha},
  journal={Nano Lett.},
  volume={25},
  number={36},
  pages={13697--13702},
  doi={10.1021/acs.nanolett.5c03764},
  year={2025},
  publisher={ACS Publications}
}

@article{verzijl13,
    author = {Verzijl, C. J. O. and Seldenthuis, J. S. and Thijssen, J. M.},
    title = {Applicability of the wide-band limit in DFT-based molecular transport calculations},
    journal = {J. Chem. Phys.},
    volume = {138},
    number = {9},
    pages = {094102},
    year = {2013},
    month = {03},
    issn = {0021-9606},
    doi = {10.1063/1.4793259},
}

@article{makri99,
  title={The linear response approximation and its lowest order corrections: An influence functional approach},
  author={Makri, Nancy},
  journal={J. Phys. Chem. B},
  volume={103},
  number={15},
  pages={2823--2829},
  year={1999},
  doi={10.1021/jp9847540},
  publisher={ACS Publications}
}

@article{walters17,
author = {Walters, Peter L. and Allen, Thomas C. and Makri, Nancy},
title = {Direct determination of discrete harmonic bath parameters from molecular dynamics simulations},
journal = {J. Comput. Chem.},
volume = {38},
number = {2},
pages = {110-115},
doi = {https://doi.org/10.1002/jcc.24527},
year = {2017}
}

@article{rokaj18,
doi = {10.1088/1361-6455/aa9c99},
url = {https://doi.org/10.1088/1361-6455/aa9c99},
year = {2018},
month = {jan},
publisher = {IOP Publishing},
volume = {51},
number = {3},
pages = {034005},
author = {Rokaj, Vasil and Welakuh, Davis M and Ruggenthaler, Michael and Rubio, Angel},
title = {Light–matter interaction in the long-wavelength limit: no ground-state without dipole self-energy},
journal = {J. Phys. B}
}

@article{dolezal24,
  title={Single-molecule time-resolved spectroscopy in a tunable STM nanocavity},
  author={Dolezal, Jiri and Sagwal, Amandeep and de Campos Ferreira, Rodrigo Cezar and Svec, Martin},
  journal={Nano Lett.},
  volume={24},
  number={5},
  pages={1629--1634},
  year={2024},
  doi={acs.nanolett.3c04314},
  publisher={ACS Publications}
}

@article{aradhya13,
  title={Single-molecule junctions beyond electronic transport},
  author={Aradhya, Sriharsha V and Venkataraman, Latha},
  journal={Nature Nanotech.},
  volume={8},
  number={6},
  pages={399--410},
  year={2013},
  doi={10.1038/nnano.2013.91},
  publisher={Nature Publishing Group UK London}
}

@article{li23,
author = {Li, Tianming and Bandari, Vineeth Kumar and Schmidt, Oliver G.},
title = {Molecular Electronics: Creating and Bridging Molecular Junctions and Promoting Its Commercialization},
journal = {Adv. Mater.},
volume = {35},
number = {22},
pages = {2209088},
doi = {https://doi.org/10.1002/adma.202209088},
year = {2023}
}

@article{Manzano20,
    author = {Manzano, Daniel},
    title = {A short introduction to the Lindblad master equation},
    journal = {AIP Advances},
    volume = {10},
    number = {2},
    pages = {025106},
    year = {2020},
    month = {02},
    issn = {2158-3226},
    doi = {10.1063/1.5115323},
    url = {https://doi.org/10.1063/1.5115323},
}

@book{BreuerPetruccioneOpen,
    author = {Breuer, Heinz-Peter and Petruccione, Francesco},
    title = {The Theory of Open Quantum Systems},
    publisher = {Oxford University Press},
    year = {2007},
    month = {01},
    isbn = {9780199213900},
    doi = {10.1093/acprof:oso/9780199213900.001.0001},
    url = {https://doi.org/10.1093/acprof:oso/9780199213900.001.0001},
}

@article{QuantumToolboxJulia,
  title = {Quantum{T}oolbox.jl: {A}n efficient {J}ulia framework for simulating open quantum systems},
  author = {Mercurio, Alberto and Huang, Yi-Te and Cai, Li-Xun and Chen, Yueh-Nan and Savona, Vincenzo and Nori, Franco},
  journal = {{Quantum}},
  issn = {2521-327X},
  publisher = {{Verein zur F{\"{o}}rderung des Open Access Publizierens in den Quantenwissenschaften}},
  volume = {9},
  pages = {1866},
  month = sep,
  year = {2025},
  doi = {10.22331/q-2025-09-29-1866},
  url = {https://doi.org/10.22331/q-2025-09-29-1866}
}

@article{nitzan02,
    author = {Nitzan, Abraham and Galperin, Michael and Ingold, Gert-Ludwig and Grabert, Hermann},
    title = {On the electrostatic potential profile in biased molecular wires},
    journal = {J. Chem. Phys.},
    volume = {117},
    number = {23},
    pages = {10837-10841},
    year = {2002},
    month = {12},
    issn = {0021-9606},
    doi = {10.1063/1.1522406},
}

@article{pleutin03,
    author = {Pleutin, Stéphane and Grabert, Hermann and Ingold, Gert-Ludwig and Nitzan, Abraham},
    title = {The electrostatic potential profile along a biased molecular wire: A model quantum-mechanical calculation},
    journal = {J. Chem. Phys.},
    volume = {118},
    number = {8},
    pages = {3756-3763},
    year = {2003},
    month = {02},
    issn = {0021-9606},
    doi = {10.1063/1.1539863},
}

@article{liu17,
  title={Voltage dependence of molecule--electrode coupling in biased molecular junctions},
  author={Liu, Zhen-Fei and Neaton, Jeffrey B},
  journal={J. Phys. Chem. C},
  volume={121},
  number={39},
  pages={21136--21144},
  year={2017},
  doi={10.1021/acs.jpcc.7b05567},
  publisher={ACS Publications}
}

@article{shen25,
  title={Ground-state charge transfer in single-molecule junctions covalent organic frameworks for boosting photocatalytic hydrogen evolution},
  author={Shen, Rongchen and Huang, Can and Hao, Lei and Liang, Guijie and Zhang, Peng and Yue, Qiang and Li, Xin},
  journal={Nat. Commun.},
  volume={16},
  number={1},
  pages={2457},
  year={2025},
  doi={10.1038/s41467-025-57662-4},
  publisher={Nature Publishing Group UK London}
}

@article{tang23,
  title={Voltage-driven control of single-molecule keto-enol equilibrium in a two-terminal junction system},
  author={Tang, Chun and Stuyver, Thijs and Lu, Taige and Liu, Junyang and Ye, Yiling and Gao, Tengyang and Lin, Luchun and Zheng, Jueting and Liu, Wenqing and Shi, Jia and others},
  journal={Nat. Commun.},
  volume={14},
  number={1},
  pages={3657},
  year={2023},
  doi={10.1038/s41467-023-39198-7},
  publisher={Nature Publishing Group UK London}
}

@article{sandik25,
author="Sandik, Gal
and Feist, Johannes
and Garc{\'{\i}}a-Vidal, Francisco J.
and Schwartz, Tal",
title="Cavity-enhanced energy transport in molecular systems",
journal="Nat. Mater.",
year="2025",
month="Mar",
day="01",
volume="24",
number="3",
pages="344--355",
issn="1476-4660",
doi="10.1038/s41563-024-01962-5"
}

@article{mondal24,
    author = {Mondal, Subhadip and Keshavamurthy, Srihari},
    title = {Cavity induced modulation of intramolecular vibrational energy flow pathways},
    journal = {J. Chem. Phys.},
    volume = {161},
    number = {19},
    pages = {194302},
    year = {2024},
    month = {11},
    issn = {0021-9606},
    doi = {10.1063/5.0236437},
}

@article{muniain24,
  title = {Unified Treatment of Light Emission by Inelastic Tunneling: Interaction of Electrons and Photons beyond the Gap},
  author = {Muniain, Unai and Esteban, Ruben and Aizpurua, Javier and Greffet, Jean-Jacques},
  journal = {Phys. Rev. X},
  volume = {14},
  issue = {2},
  pages = {021017},
  numpages = {29},
  year = {2024},
  month = {Apr},
  publisher = {American Physical Society},
  doi = {10.1103/PhysRevX.14.021017}
}

@article{zhang15,
    author="Zhang, Peng",
    title="Scaling for quantum tunneling current in nano- and subnano-scale plasmonic junctions",
    journal="Sci. Rep.",
    year="2015",
    month="May",
    day="19",
    volume="5",
    number="1",
    pages="9826",
    issn="2045-2322",
    doi="10.1038/srep09826"
}

@article{fay23,
    author = {Fay, Thomas P. and Limmer, David T.},
    title = {Spin selective charge recombination in chiral donor–bridge–acceptor triads},
    journal = {J. Chem. Phys.},
    volume = {158},
    number = {19},
    pages = {194101},
    year = {2023},
    month = {05},
    issn = {0021-9606},
    doi = {10.1063/5.0150269},
}

@article{fay21,
  title={Origin of chirality induced spin selectivity in photoinduced electron transfer},
  author={Fay, Thomas P and Limmer, David T},
  journal={Nano Lett.},
  volume={21},
  number={15},
  pages={6696--6702},
  year={2021},
  doi={10.1021/acs.nanolett.1c02370},
  publisher={ACS Publications}
}

@article{chen16,
    author ="Chen, Lipeng and Gelin, Maxim F. and Chernyak, Vladimir Y. and Domcke, Wolfgang and Zhao, Yang",
    title  ="Dissipative dynamics at conical intersections: simulations with the hierarchy equations of motion method",
    journal  ="Faraday Discuss.",
    year  ="2016",
    volume  ="194",
    issue  ="0",
    pages  ="61-80",
    publisher  ="The Royal Society of Chemistry",
    doi  ="10.1039/C6FD00088F"
}

@article{marcus85,
title = {Electron transfers in chemistry and biology},
journal = {Biochim. Biophys. Acta},
volume = {811},
number = {3},
pages = {265-322},
year = {1985},
issn = {0304-4173},
doi = {https://doi.org/10.1016/0304-4173(85)90014-X},
url ={https://www.sciencedirect.com/science/article/pii/030441738590014X},
author = {R.A. Marcus and Norman Sutin}
}

\end{document}


\renewcommand{\thesection}{S\arabic{section}}
\renewcommand{\thesubsection}{S\arabic{section}\Alph{subsection}}
\renewcommand{\theequation}{S\arabic{equation}}
\renewcommand{\thefigure}{S\arabic{figure}}

\title{Supporting Information: Steady--State Current Signatures of Strong Light--Matter Coupling in Single--Molecule Junctions}
\author{Kritanjan Polley}
\affiliation{Simons Center for Computational Physical Chemistry at New York University, New York, New York 10003, USA}
\author{Norah M. Hoffmann}
\affiliation{Simons Center for Computational Physical Chemistry at New York University, New York, New York 10003, USA}
\affiliation{Department of Chemistry, New York University, New York, New York 10003, USA}
\affiliation{Department of Physics, New York University, New York, New York 10003, USA}
\email{nmh6061@nyu.edu}

\maketitle

\section{Mapping Approaches}\label{secMapping}

\subsection{Li-Miller Mapping}
The Li-Miller (LM)~\cite{li12} mapping is based on expressing each fermionic degree of freedom as a two-state system, characterized by the Pauli spin matrices ($\bm{\kappa}_x$, $\bm{\kappa}_y$, and $\bm{\kappa}_z$), and the isomorphism between the Pauli matrices and quaternions,
\begin{equation}
    \bm{I} \rightarrow \bm{1}, \, -i\bm{\kappa}_x \rightarrow \hat{\bm{i}}, \, -i\bm{\kappa}_y \rightarrow \hat{\bm{j}}, \, -i\bm{\kappa}_z \rightarrow \hat{\bm{k}},
\end{equation}
and the anti-commutative relations of the quaternions
\begin{gather}
    \hat{\bm{i}}\hat{\bm{j}} = -\hat{\bm{j}}\hat{\bm{i}} = \hat{\bm{k}}, \, \hat{\bm{j}}\hat{\bm{k}} = -\hat{\bm{k}}\hat{\bm{j}} = \hat{\bm{i}}, \, \hat{\bm{k}}\hat{\bm{i}} = -\hat{\bm{i}}\hat{\bm{k}} = \hat{\bm{j}}, \hat{\bm{i}}\hat{\bm{i}} = \hat{\bm{j}}\hat{\bm{j}} = \hat{\bm{k}}\hat{\bm{k}} = -\bm{1},
\end{gather}
where $i=\sqrt{-1}$ is the imaginary unit while $\hat{\bm{i}}, \hat{\bm{j}}, \hat{\bm{k}}$ are quaternions. The creation and annihilation operators can be expressed as
\begin{gather}
 c^{\dagger} = \frac{1}{2}(\bm{\kappa}_x + i\bm{\kappa}_y) = \frac{1}{2} (i\hat{\bm{i}}-\hat{\bm{j}}), \quad c = \frac{1}{2}(\bm{\kappa}_x - i\bm{\kappa}_y) = \frac{1}{2} (i\hat{\bm{i}}+\hat{\bm{j}}).
\end{gather}
The quaternions are mapped into a 2D Cartesian space,
\begin{gather}
    \frac{i}{\sqrt{2}}\hat{\bm{i}} \rightarrow \begin{pmatrix}
        x \\y
    \end{pmatrix}, \, \frac{1}{\sqrt{2}}\hat{\bm{j}} \rightarrow \begin{pmatrix}
        p_x \\ p_y
    \end{pmatrix}, \, \frac{i}{2}\hat{\bm{i}}_j\hat{\bm{j}}_k = \begin{pmatrix}x_j \\y_j \end{pmatrix} \times \begin{pmatrix} p_{x,k} \\ p_{y,k} \end{pmatrix} = x_jp_{y,k}-y_jp_{x,k}.
\end{gather}
The products of creation-annihilation operators take the form
\begin{align}
    c_{m}^{\dagger} c_{n} + c_{n}^{\dagger} c_{m} & = x_{m}p_{y,n} - y_{m}p_{x,n} + x_{n}p_{y,m} - y_{n}p_{x,m}, \label{eqCmnSumLM} \\
    c_{m}^{\dagger} c_{n} - c_{n}^{\dagger} c_{m} & = x_{m}y_{n} -x_{n}y_{m} + p_{x,n}p_{y,m} - p_{x,m}p_{y,n} \label{eqCmnDifLM}, 
\end{align}

\subsection{Complete Quasiclassical Mapping}
The complete quasiclassical map (CQM)~\cite{levy19} replaces the Li-Miller map with
\begin{equation}
    \sqrt{\frac{i}{2}} \hat{\bm{i}} \rightarrow \begin{pmatrix} x\\ p_x \end{pmatrix}, \quad \sqrt{\frac{i}{2}} \hat{\bm{j}} \rightarrow \begin{pmatrix} y\\ p_y \end{pmatrix},
\end{equation}
and this change modifies the products of creation-annihilation operators as follows
\begin{align}
    c_{m}^{\dagger} c_{n} + c_{n}^{\dagger} c_{m} & = x_{m}p_{y,n} - y_{m}p_{x,n} + x_{n}p_{y,m} - y_{n}p_{x,m}, \label{eqCmnSumCQ}\\
    c_{m}^{\dagger} c_{n} - c_{n}^{\dagger} c_{m}  & = i(x_{m}p_{x,n} - x_{n}p_{x,m} +y_{m}p_{y,n} -y_{n}p_{y,m} ).\label{eqCmnDifCQ}
\end{align}

While both the Li-Miller and Complete Quasiclassical Mapping (CQM) provide frameworks for representing fermionic degrees of freedom in a phase-space representation, they differ in their treatment of the underlying algebra. Although the independent electron approximation correctly captures the time-dependence of one-body operators, it requires one to change important details in the way that one calculates observables for many-body problems.

\subsection{Bosonic Mapping Approaches}
Bosonic mapping approaches are accurate for non-interacting bosons, and therefore, they yield the correct dynamics for non-interacting fermions.~\cite{sun21} While the Meyer–Miller mapping of fermions is not exact for interacting fermions or when coupled to bosonic degrees of freedom, it can still provide an accurate description.~\cite{jung23} There are several other works based on exact bosonization and cartesian mapping of fermions.~\cite{langmann15,montoya23,dhar06,chen20}

\section{Mapping Hamiltonian}\label{appMappingHamil}
In the limit where the diabatic couplings are real, \textit{i.e.}, when the Hamiltonian lacks terms of the form $c_{m}^{\dagger} c_{n} - c_{n}^{\dagger} c_{m}$, the Li-Miller and CQM approaches yield identical results. The mapping Hamiltonian has the form
\begin{align}
    H_s & =  \sum_{j} h_{jj} (\bm{R})\left(x_{j}p_{y,j} - y_{j}p_{x,j}\right)  + \sum_{j>k} h_{jk}(\bm{R})(x_{j}p_{y,k} - y_{j}p_{x,k} + x_{k}p_{y,j} - y_{k}p_{x,j}), \\
    H_{b} & =  \sum_{g,l} \epsilon_{g,l} \left(\bar{x}_{g,l}\bar{p}_{y,gl} - \bar{y}_{g,l}\bar{p}_{x,gl}\right), \quad H_{sb} =  \sum_{j,l,g} t_{jgl} \Big[\left(\bar{x}_{g,l}p_{y,j} - \bar{y}_{g,l}p_{x,j}\right) + \left(x_{j}\bar{p}_{y,gl} - y_{j}\bar{p}_{x,gl}\right)\Big],\\
    H_{p} & =  \frac{\hbar\omega_{\mathrm{cav}}}{2} (p_{\mathrm{cav}}^2 + q_{\mathrm{cav}}^2), \quad H_{sp} =  \sum_{j>k} \sqrt{2}\lambda_{jk}(\bm{R})q_{\mathrm{cav}}(x_{j}p_{y,k} - y_{j}p_{x,k} + x_{k}p_{y,j} - y_{k}p_{x,j}),
\end{align}
where $x_j$, $y_j$, $p_{x,j}$, and $p_{y,j}$ are phase space variables for the system degrees of freedom, and the variables with overbar are for the bath degrees of freedom. The current can be calculated by averaging over the trajectories to determine the rate of change in population of the electrode modes,
\begin{align}
    I_{l} (t) = & - \sum_{g} \frac{d}{dt} \Big\langle  d^{\dagger}_{g,l}d_{g,l} \Big\rangle = - \sum_{g} \frac{d}{dt} \Big\langle \bar{x}_{g,l}\bar{p}_{y,gl} - \bar{y}_{g,l}\bar{p}_{x,gl} \big\rangle, \\
    = & - \sum_{j,g} \Big\langle t_{jgl} \big(\bar{x}_{g,l}p_{x,j} + \bar{y}_{g,l}p_{y,j} - x_{j}\bar{p}_{x,gl} - y_{j}\bar{p}_{y,gl}\big)\Big\rangle,
\end{align}
and $I_{\text{tot}} (t) = \frac{1}{2} (I_{L} - I_{R})$.
The time evolution of the phase-space variables is governed by Hamilton's equations of motion,
\begin{align}
    \dot{x}_j & = \frac{\partial H}{\partial p_{x,j}}, \dot{y}_j = \frac{\partial H}{\partial p_{y,j}}, \dot{p}_{x,j}  = -\frac{\partial H}{\partial x_{j}}, \dot{p}_{y,j} = -\frac{\partial H}{\partial y_{j}}, \label{eqxpdot}
\end{align}
and similar for bath degrees of freedom.

\section{Initial Condition for Mapping Variables}\label{appInitialCond}
The initial conditions for the system and electrode modes are sampled as per Ref.~\citenum{li13}. A brief description of this process is provided below. To enforce the quantum statistics for each mode, the initial conditions are sampled from a Fermi-Dirac distribution. This selection of initial condition guarantees the precise recovery of the correct statistical behavior at $t=0$.
\begin{align}
    x_{k} & = r_{k} \cos \theta_{k},\quad p_{x,k} = p_{r,k} \cos \theta_{k} -n_{k} \frac{\sin \theta_{k}}{r_{k}}, \\
    y_{k} & = r_{k} \sin \theta_{k},\quad p_{y,k} = p_{r,k} \sin \theta_{k} +n_{k} \frac{\cos \theta_{k}}{r_{k}},
\end{align}
where 
\begin{equation}
    n_{k} = \begin{cases}
        0 & \xi_{k} > \left( 1+ e^{\beta (\epsilon_{k}-\mu_{L/R})}\right)^{-1}\\
        1 & \xi_{k} \le \left( 1+ e^{\beta (\epsilon_{k}-\mu_{L/R})}\right)^{-1}
    \end{cases},
\end{equation}
with $\xi_{k}\in [0,1]$ and $\theta_k \in [0,2\pi]$ are random numbers, and $\beta$ is the inverse thermal energy. While the specific selection of the radial components $p_{r,k}$ and $r_{k}$ is not unique, the mapping remains physically valid provided the initial population is given by $n_{k}=x_{k}p_{y,k}-y_{k}p_{x,k}$. We adopt the convention $r_{k}=1$ and $p_{r,k}=0$.~\cite{li13,li14} The system degrees of freedom are similarly sampled to ensure that the orbitals are initially in a vacuum state, \textit{i.e.}, $n_k=0$.

The hopping rates between electrodes and the fermionic energy levels are obtained using
\begin{equation}
    t_{jgl} = \sqrt{\frac{J_{\mathrm{fermion}}(\epsilon_{jgl})\Delta \epsilon}{2\pi}},
\end{equation}
where $\Delta \epsilon = 2\epsilon_{\mathrm{max}}/(G-1)$ is the discretization of the electrode energy levels, $G$ is the number of modes in the electrode $l$ and $2\epsilon_{\mathrm{max}}$ is the energy range in the electrode. The form of the spectral density, $J_{\mathrm{fermion}}(\epsilon)$, is given in Eq.~\eqref{eqSpecDen} in the main text.

\begin{figure}
    \centering
    \includegraphics[width=0.9\linewidth]{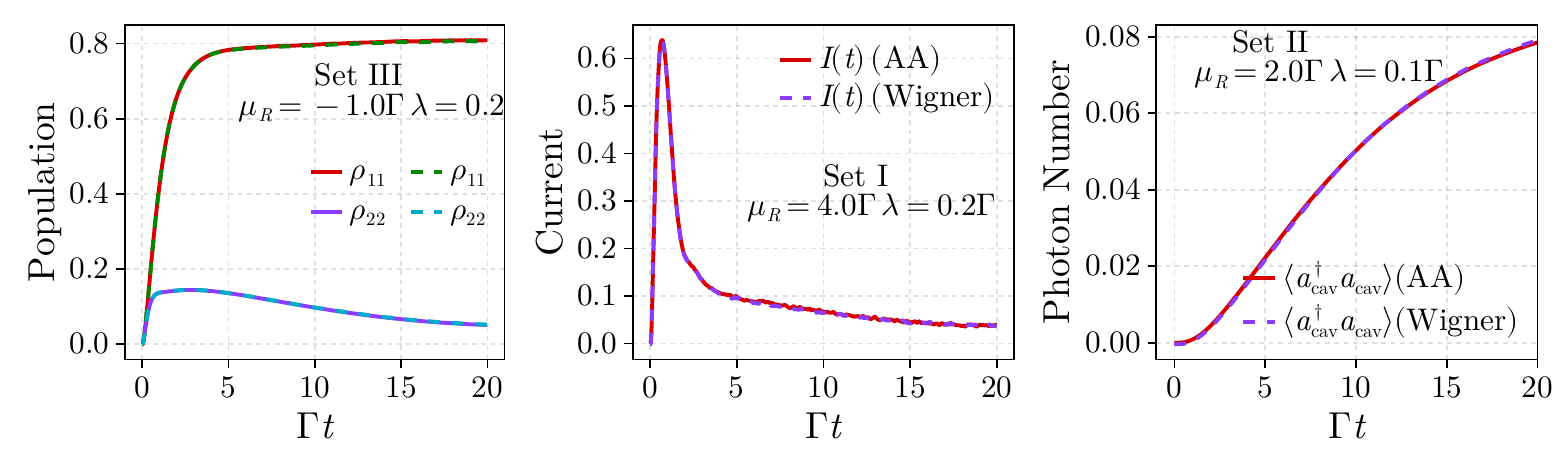}
    \caption{Two distinct sampling methods for photon mode, action-angle variables and Wigner sampling, yield essentially same results. The leftmost panel illustrates the propagation of the electronic population, while the middle panel depicts the propagation of the total current. The rightmost panel shows the photon number as a function of time. The agreement between the two methods confirms the robustness of the sampling procedure. Specific parameters are provided within each panel.}
    \label{figCompareSampling}
\end{figure}

The photon and vibrational modes are sampled from a Wigner transformed Boltzmann distribution at absolute zero temperature, or from a phase space of constant action with a random phase. The Wigner transformed Boltzmann distribution at 0K is given by
\begin{equation}
    \rho(p_{\sigma},q_{\sigma}) \propto \exp \{ - (p_{\sigma}^2 + q_{\sigma}^2) \}, \quad {\sigma}=t,c,\mathrm{cav}, \label{eqInitialWigner}
\end{equation}
while in the action-angle formulation, the phase space variables are sampled as
\begin{align}
    p_{\sigma} = & \sqrt{2(n_{\sigma} + \gamma_{vib})} \cos \theta_{\sigma}, \quad q_{\sigma} = -\sqrt{2(n_{\sigma} + \gamma_{vib})} \sin \theta_{\sigma}, \label{eqInitialQ}
\end{align}
where $\theta_{\sigma} \in [0,2\pi]$, $n_{\sigma}$ is the initial population at $t=0$, and $\gamma_{vib}$ is the zero point energy for the vibrational modes. We have used $\gamma_{vib}=1/2$ in these calculations. The choice of sampling method for the photon mode does not significantly impact the observables, as demonstrated in Fig.~\ref{figCompareSampling}.

The initial conditions for the bosonic bath oscillators were chosen from a Wigner-transformed Boltzmann distribution,
\begin{align}
    \rho (\bm{P}, \bm{Q}) & = \prod \limits_{\alpha=1} \frac{\xi_{\alpha}}{\pi\hbar} \exp \left[ \frac{-2\xi_{\alpha}}{ \hbar \omega_{\alpha}}\left( \frac{P_{\alpha}^{2}}{2m_{\alpha}} +\frac{m_{\alpha}\omega_{\alpha}^2 Q_{\alpha}^2}{2} \right) \right], \label{eqWigner}
\end{align}
where $\xi_{\alpha} = \tanh (\beta \hbar \omega_{\alpha} /2)$. The bath modes are sampled as~\cite{makri99,walters17}
\begin{align}
    J_{\mathrm{boson}}(\omega) & = \frac{\pi}{2} \sum_{\alpha}^N \frac{c_{\alpha}^2}{m_{\alpha}\omega_{\alpha}}\delta (\omega -\omega_{\alpha}) \simeq 2\eta \frac{\omega \omega_c}{\omega^2 + \omega_c^2}, \label{eqJw} \\
    \omega_{\alpha} & = \omega_c \tan \left(\frac{\alpha}{N}\mathrm{tan}^{-1} \left( \frac{\omega_N}{\omega_c} \right) \right), \quad c_{\alpha} = \omega_{\alpha}\sqrt{\frac{2\eta m_{\alpha}}{N}}. \label{eqca}
\end{align}

\section{Propagation Details}\label{secNumericalDetails}
500 fermionic levels for each electrode are employed to model the fermionic environment. The equations are simulated using a RK4 algorithm with a time step of $0.005\,\Gamma t$, averaging over $5\times 10^6$ trajectories. For bosonic bath mode discretizations, we have used $\omega_N=15\omega_c$, and $N=100$.

\section{Bare Fermionic 2LS in a Cavity}\label{appBare2LS}

We investigate a system with two quantum spinless dots between two electrodes, coupled to a photon cavity. The Hamiltonian for this system is given by, 
\begin{align}
    H_s & = \epsilon_1 c_1^{\dagger}c_1 + \epsilon_2 c_2^{\dagger}c_2 \label{eqAppHS}\\
    H_b & = \sum_{l=L,R}\sum_g \epsilon_{g,l}d_{g,l}^{\dagger}d_{g,l}, \quad H_{sb} = \sum_{j=1,2}\sum_{g} \sum_{l=L,R} t_{j,g,l} c_{j}^{\dagger}d_{g,l} + \text{h.c.},\\
    H_p & = \hbar \omega_{\mathrm{cav}} \bigg(a^{\dagger}_{\mathrm{cav}}a_{\mathrm{cav}} + \frac{1}
    {2}\bigg), \quad H_{ps} = \lambda (a_{\mathrm{cav}}+a^{\dagger}_{\mathrm{cav}})(c_1^{\dagger}c_2+c_2^{\dagger}c_1). \label{eqAppHPS}
\end{align}

\begin{figure}
    \centering
    \includegraphics[width=\linewidth]{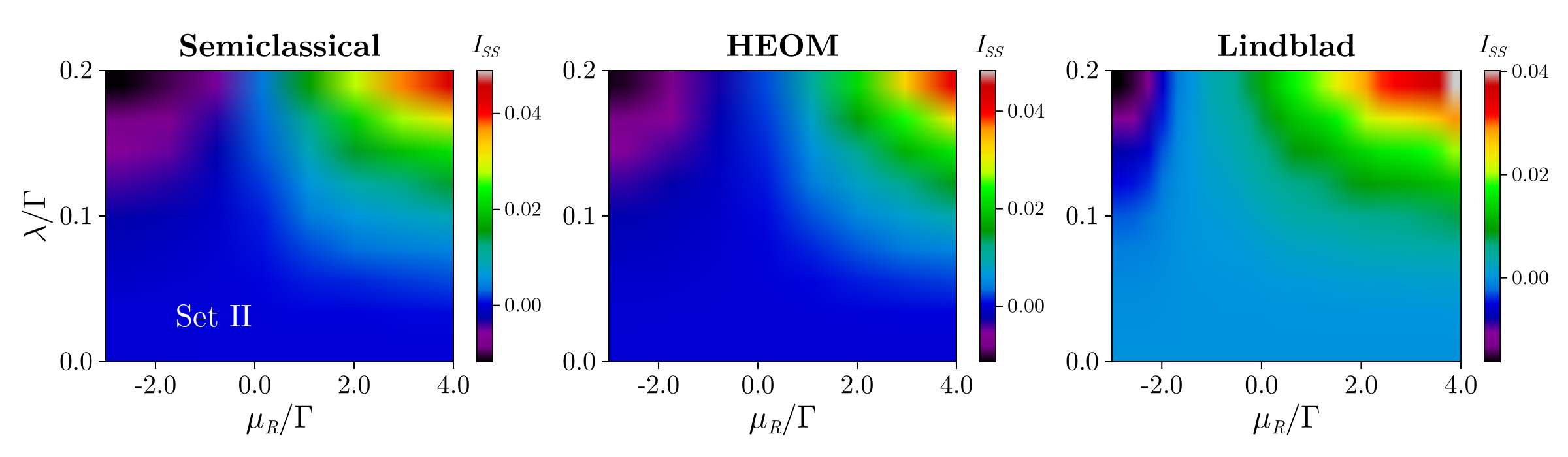}
    \caption{Heatmaps display the current as a function of bias and light-matter coupling for Set II parameters (\textit{c.f.} Table.~\ref{tabParams} in the main text), for different values of light matter coupling ($\lambda$). The left panel shows results obtained via the semiclassical mapping approach, the middle panel presents the numerically exact HEOM results, and the right panel shows the Lindblad master equation results.  The theoretical framework for the master equation approach is detailed in Sec.~\ref{secLindblad}. The current expressed in the units of $e\hbar/\Gamma$.}
    \label{figSteadyStateSet2Heat}
\end{figure}

The steady-state current, calculated using Set II parameters from Table~\ref{tabParams} in the main text, is depicted in Fig.~\ref{figSteadyStateSet2Heat}. At zero bias ($\mu_L=\mu_R=0$), the steady-state current vanishes, a result consistent with our semiclassical mapping framework. In the regime of negative $\mu_R$ (or, positive $\mu_L$), as observed on the left side of the heatmaps, the current is significantly suppressed due to the lack of spectral alignment between the electrode chemical potentials and the electronic energy levels of the junction. Conversely, for $\mu_R>0$, the steady-state current gradually increases, reaching a maximum when the electronic energy gap aligns with the bias voltage for a given light-matter coupling. In the present setup, the light-matter coupling acts as the exclusive mechanism for electronic population transfer between electrodes, as evidenced by the absence of current at zero $\lambda$.

\section{Two dimensional potential}\label{secSIpotential}
We employ a two-dimensional oscillator ($\bm{R}\equiv \{Q_c,Q_t\}$) for the nuclear potential,~\cite{chen16,gu20a,gu20b,cho22} which takes the form
\begin{align}
    H_s & = \sum_{j=1,2}h_{jj} c_j^{\dagger}c_j + \gamma Q_c (c_1^{\dagger}c_2 + c_2^{\dagger}c_1), \label{eqPotFull} \\ 
    h_{jj} & = h_0 + \epsilon_j + \kappa_j Q_t, \, h_0 = \sum_{\sigma=t,c} \frac{\hbar \omega_{\sigma}}{2} \big( Q_{\sigma}^2 + P_{\sigma}^2\big), \label{eqPoth0}
\end{align}
where $\gamma$ is the diabatic bilinear coupling strength between the fermionic modes and the $Q_c$ mode. We set the value of the diabatic coupling parameter, $\gamma$, to $0.05\Delta$. For other parameters of the nuclear potential, we use $\hbar\omega_c=0.2\Delta$, $\hbar\omega_t=0.1\Delta$, $\kappa_1=-0.2\Delta$, and $\kappa_2=0.3\Delta$. The shape of the potential is shown in panel (A) of Fig.~\ref{figSteadyStateLambda1} in the main text. We used Meyer-Miller-Stock-Thoss~\cite{meyer79a,meyer79b,stock97,thoss99} mapping for the bosonic nuclear modes.

\section{Lindblad Master Equation Approach}\label{secLindblad}
To facilitate a comparison with the semiclassical mapping results, we describe the trace-preserving and completely positive evolution of the reduced density matrix, ($\rho_{\mathrm{sys}} = \mathrm{Tr}_{\mathrm{bath}}[\rho_{\mathrm{tot}}]$) using the Lindblad master equation,
\begin{align}
    \frac{d}{dt} \rho_{\mathrm{sys}} (t) & = -\frac{i}{\hbar} [H_{\mathrm{sys}}, \rho_{\mathrm{sys}}] + \sum_{k} C_k \rho_{\mathrm{sys}} C_k^{\dagger} - \frac{1}{2}C_k^{\dagger} C_k \rho_{\mathrm{sys}} - \frac{1}{2} \rho_{\mathrm{sys}} C_k^{\dagger} C_k,
\end{align}
where the operators $C_k$ are usually referred as jump operators, standard notation follows that of widely used literature.~\cite{Manzano20,BreuerPetruccioneOpen} We used \texttt{QuantumToolBox} library in \texttt{Julia} to propagate the master equation.~\cite{QuantumToolboxJulia}

The fundamental physical complexity of modeling open quantum systems necessitates several crucial approximations to derive a computationally feasible master equation. This framework relies on the assumption of weak coupling between the system and the bath, which allows us to treat the total density matrix as separable  ( $\rho_{\mathrm{tot}}\approx \rho_{\mathrm{sys}} \otimes \rho_{\mathrm{bath}}$ ) and the state of the bath remains nearly constant, and the system and bath remain separable throughout the evolution. Furthermore, the Markovian approximation is applied to account for instantaneous environmental memory loss. Finally, the secular approximation neglects rapidly oscillating terms in the interaction picture to simplify the effective Hamiltonian.

\begin{figure}
    \centering
    \includegraphics[width=0.8\linewidth]{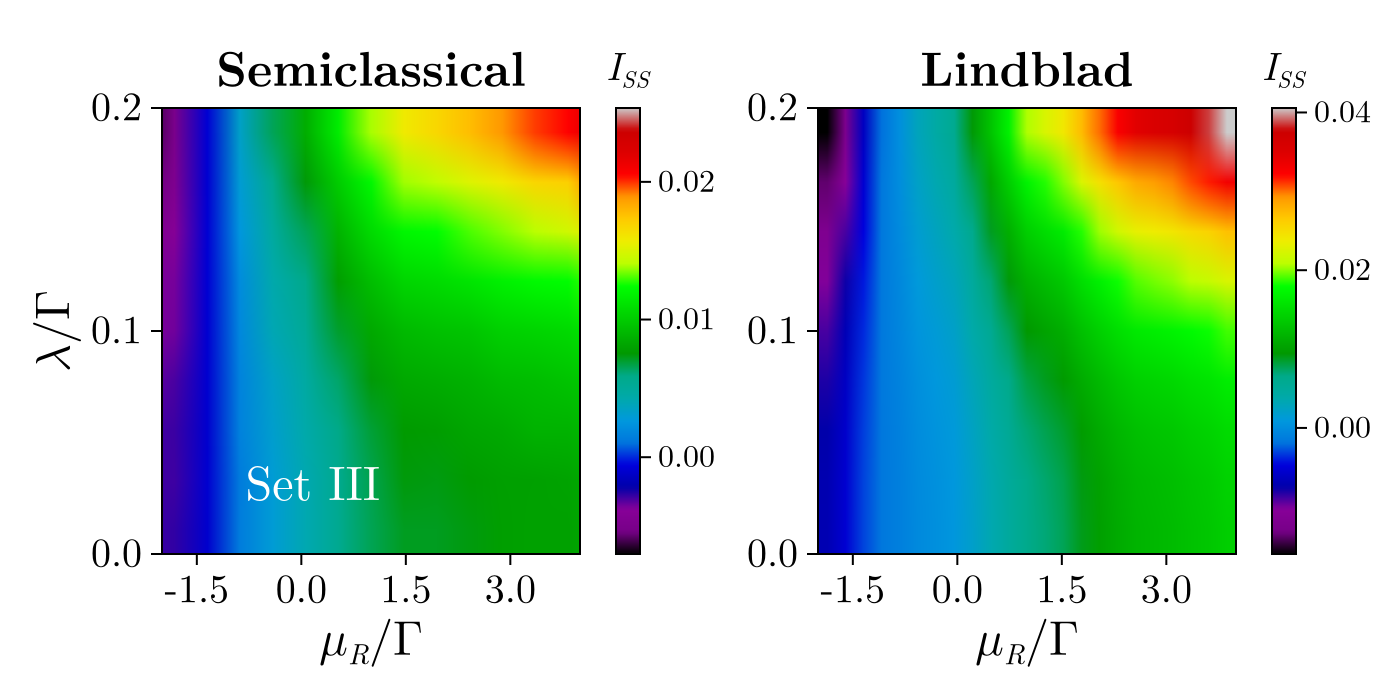}
    \caption{Results for Set III parameters under varying light-matter coupling on a STM-BJ setup are compared between the Lindblad master equation approach and the semiclassical mapping approach. The qualitative agreement validates the semiclassical framework in the presence of nuclear degrees of freedom.}
    \label{figMasterEqCompare}
\end{figure}

The impact of the approximations inherent in the Lindblad framework is evident in the rightmost panel of Fig.~\ref{figSteadyStateSet2Heat}. While the master equation approach successfully captures the qualitative trends observed in the numerically exact HEOM method, there is a measurable discrepancy in the absolute magnitude of the steady-state current. This is primarily attributed to the different treatment of the electrode coupling; the Lindblad approach employs a Fermi-Dirac distribution for the hopping rates, whereas the semiclassical approach utilizes a wide-band limit approximation. When nuclear modes are included, the results remain qualitatively consistent, as shown in Fig.~\ref{figMasterEqCompare}. 

A primary limitation of the Lindblad master equation in this context arises from the assumption of weak system-bath coupling. In the STM-BJ setup, the interaction between the molecule and the electrodes is inherently strong, which may challenge the validity of the separable total density matrix approximation. Furthermore, the Markovian approximation assumes instantaneous environmental memory loss; however, in driven systems like the STM-BJ, the bath correlation time can be sufficiently long to retain significant information about previous system dynamics. These factors contribute to the observed discrepancies in the magnitude of the steady-state current compared to the numerically exact HEOM and semiclassical results.

\section{Gauge Transformation with 2LS}\label{secGauge}

\begin{figure}
    \centering
    \includegraphics[width=0.8\linewidth]{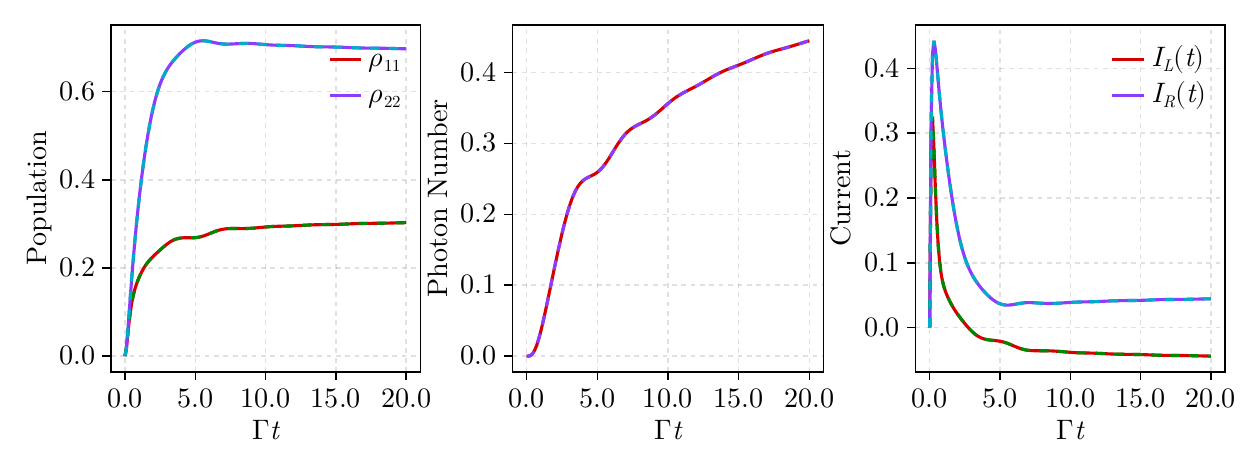}
    \caption{A comparative analysis of length (solid lines) and velocity (dash lines) gauges for various observables is presented in the figure above. The parameters utilized in this analysis are $\epsilon_1-\epsilon_2=\Gamma$, $\omega_{\mathrm{cav}}=\Gamma$, $\lambda=0.5\Gamma$, $W=10\Gamma$, $\mu_R-\mu_L=2\Gamma$, and $k_BT=\Gamma$. The left panel shows the population of the two states, initially devoid of any particles, and the total photon number on the cavity mode is illustrated on the middle panel, respectively. The rightmost panel showcases the current flowing to the left and right electrodes. Notably, the unitary transformation does not influence the dynamics of the system.}
    \label{figGauge}
\end{figure}

To transform from velocity gauge to length gauge, we use the following Power-Zienau-Woolley transformation matrix, a unitary operator,~\cite{babiker83,woolley20}
\begin{align}
    U & \equiv \exp \bigg \{ -i\frac{\lambda}{\hbar\omega_{\mathrm{cav}}} (c_1^{\dagger}c_2 + c_2^{\dagger}c_1) (a_{\mathrm{cav}} + a^{\dagger}_{\mathrm{cav}}) \bigg\},
\end{align}
and the transformed Hamiltonian can be obtained as $\bar{H} = U H U^{\dagger}$ where $H$ is described in Eqs.~\eqref{eqAppHS}-\eqref{eqAppHPS}. The transformed Hamiltonian will have the form
\begin{align}
    \bar{H} & = \frac{\epsilon_1 + \epsilon_2}{2} \sigma_N + \frac{\epsilon_1 - \epsilon_2}{2} (\sigma_z \cos (2\phi) - \sigma_y \sin(2\phi)) + \sum_{l=L,R}\sum_g \epsilon_{g,l}d_{g,l}^{\dagger}d_{g,l} \nonumber \\
    & + \sum_{g,l} \Big [ (t_{1gl} \cos (\phi) -it_{2gl}\sin (\phi) ) c_{1}^{\dagger}d_{gl} + (t_{2gl}\cos (\phi) -it_{1gl}\sin (\phi) )c_2^{\dagger}d_{gl} + \text{h.c.} \Big]\nonumber \\
    & + \hbar \omega_{\mathrm{cav}} a^{\dagger}_{\mathrm{cav}}a_{\mathrm{cav}} + \left( \frac{\lambda^2}{\hbar\omega_{\mathrm{cav}}} + \frac{\hbar\omega_{\mathrm{cav}}}{2} \right) - i\lambda \sigma_x (a_{\mathrm{cav}} - a^{\dagger}_{\mathrm{cav}}),
\end{align}
where $\sigma_N = (c_1^{\dagger}c_1 + c_2^{\dagger}c_2)$, $\sigma_z = (c_1^{\dagger}c_1 - c_2^{\dagger}c_2)$, $\sigma_x = (c_1^{\dagger}c_2 + c_2^{\dagger}c_1)$, $\sigma_y = -i(c_1^{\dagger}c_2 - c_2^{\dagger}c_1)$, and $\phi = \frac{\lambda}{\hbar \omega_{\mathrm{cav}}} (a_{\mathrm{cav}}+ a^{\dagger}_{\mathrm{cav}})$. 

Because the transformation between the length and velocity gauges is unitary, the physical observables, including state populations and total current, remain invariant. The agreement shown in Fig.~\ref{figGauge} serves as a rigorous consistency check, confirming that the numerical evolution is independent of the choice of gauge.

\bibliography{reference}